\documentclass[%
 preprint,
groupedaddress,
 amsmath,amssymb,
 aps,
longbibliography,
superscriptaddress
]{revtex4-1}

\usepackage{graphicx}
\usepackage{dcolumn}
\usepackage{bm}
\usepackage{framed}
\usepackage{colortbl}
\usepackage{color}
\definecolor{Gray}{gray}{0.87}
\usepackage{subfigure}
\usepackage{epstopdf}
\usepackage{multirow}
\usepackage{amsmath}
\usepackage{amssymb}

\newcommand{\gO}[1]{\hbox{$\mathcal{#1}$}}
\newcommand{\braces}[2]{\hbox{$\left[\begin{array}{c} #1 \\ #2 \end{array} \right]$}}
\usepackage[mathlines]{lineno}

\newdimen\fancychapwidth
\newdimen\fancychapsep

\newcommand{\cita}[1]{\hfill\parbox{\fancychapwidth}{\em #1}	
   \vspace{\fancychapsep}\par}

\begin{document}
\title{Bi-layer voter model: Modeling intolerant/tolerant positions and bots in opinion dynamics}

\author{Didier A. Vega-Oliveros}\email{davo@unicamp.br}
\affiliation{Institute of Computing, University of Campinas, Campinas, SP, Brazil}
\affiliation{Center for Complex Networks and Systems Research, Luddy School of Informatics, Computing, and Engineering, Indiana University, Bloomington, IN, USA}

\author{Helder L. C. Grande}\email{heldercgrande@gmail.com}
\affiliation{National Institute for Space Research (INPE), São José dos Campos, SP, Brazil}

\author{Flavio Iannelli}\email{flavio.iannelli@business.uzh.ch}
\affiliation{URPP Social Networks, Universit\"at Z\"urich, Andreasstrasse 15, CH-8050 Z\"urich, Switzerland}

\author{Federico Vazquez}\email{fede.vazmin@gmail.com}
\affiliation{Instituto de Cálculo, FCEN, Universidad de Buenos Aires and CONICET, Buenos Aires, Argentina}

\begin{abstract}
The diffusion of opinions in Social Networks is a relevant process for adopting positions and attracting potential voters in political campaigns. Opinion polarization, bias, targeted diffusion, and the radicalization of postures are key elements for understanding the voting dynamics' challenges.  In particular, social bots are currently a new element that can have a pronounced effect on the formation of opinions during electoral processes by, for instance, creating fake accounts in social networks to manipulate elections.  Here, we propose a voter model incorporating bots and radical or intolerant individuals in the decision-making process.  The dynamics of the system occur in a multiplex network of interacting agents composed of two layers, one for the dynamics of opinions where agents choose between two possible alternatives, and the other for the tolerance dynamics, in which agents adopt one of two tolerance levels.  The tolerance accounts for the likelihood to change opinion in an interaction, with tolerant (intolerant) agents switching opinion with probability $1.0$ ($\gamma \le 1$).  We find that intolerance leads to a consensus of tolerant agents during an initial stage that scales as $\tau^+ \sim \gamma^{-1} \ln N$, who then reach an opinion consensus during the second stage in a time that scales as $\tau \sim N$, where $N$ is the number of agents. Therefore, very intolerant agents ($\gamma \ll 1$) could considerably slow down dynamics towards the final consensus state.  We also find that the inclusion of a fraction $\sigma_{\mathbb{B}}^-$ of bots breaks the symmetry between both opinions, driving the system to a consensus of intolerant agents with the bots' opinion.  Thus, bots eventually impose their opinion to the entire population, in a time that scales as $\tau_B^- \sim \gamma^{-1}$ for $\gamma \ll \sigma_{\mathbb{B}}^-$ and $\tau_B^- \sim 1/\sigma_{\mathbb{B}}^-$ for $\sigma_{\mathbb{B}}^- \ll \gamma$. 
\end{abstract}
\maketitle

\section{\label{sec:introduction}Introduction}

\cita{``if a society is tolerant without limit, its ability to be tolerant is eventually seized or destroyed by the intolerant.''\\[5pt]
\rightline{{\rm --- Karl Popper}}}

The voter model describes a simple process for opinion dynamics and consensus in a population of agents that can hold one of two different opinions ($A$ and $B$) \cite{Clifford-1973,liggett1975}.  In a single step of the dynamics, a voter chosen at random adopts a random neighbor's opinion. This step is repeated until voters' population eventually reaches a state of consensus in a finite system, where all agents share the same opinion.  Due to its simplicity and analytical tractability, the voter model has become a paradigmatic model to study basic properties of opinion diffusion, and the dynamics of elections \cite{castellano}.  After its introduction in two independent works, by Clifford in $1973$ \cite{Clifford-1973} and soon lately by Liggett in $1975$ \cite{liggett1975}, many extensions of the voter model have been proposed in the scientific literature to mimic more realistic or complex scenarios of social dynamics, such as considering multiple opinions \cite{vazquez_2004,volovik_2009,vazquez_2019}, heterogeneity in transition rates \cite{masuda_2010,Vega-Oliveros_2017}, and complex interaction topologies that are static \cite{suchecki_2005,sood_2008,suchecki_2005_b,sood_2005,Vega-Oliveros_2020,vazquez-2008-2} or evolve in time \cite{vazquez-2008-1,Demirel-2014}, where clusters of opposite opinions coexist.  Other works have studied how the presence of agents that never change opinion (stubborn individuals) affects the dynamics and consensus properties of the system \cite{galan_2004,galan_2007,mobila_2007}.  Moreover, the introduction of personalized information \cite{marzo}, reinforcing the political orientation of an agent when its opinion changes, has shown to prevent global consensus for strong captured information change, showing the phenomena of strengthening political positions observed in many countries.  This polarization behavior has also recently been explored through multistate voter models that include a mechanism of opinion reinforcement, which is a consequence of exchanging persuasive arguments \cite{LaRocca-2014,Velasquez-2018,vazquez-2020,Saintier-2020}.  Another implementation of the voter model has investigated the role of confidence in individuals by introducing two states per agent, its opinion, and its level of commitment to the opinion: unsure or tolerant and confident or intolerant \cite{volovik_2012}.  After interacting with an agent of the opposite opinion, a tolerant agent can change its opinion, while an intolerant agent becomes tolerant but keeps its opinion.  It is found that consensus is achieved very quickly in a mean-field setup (all-to-all interactions). At the same time, in square lattices of finite dimensions, the system reaches a metastable state where clusters of opposite opinions coexist for very long times until consensus is eventually reached.   

Given the propensity of polarization in societies and the emergence of echo chambers within political conversations in online social networks (OSN)~\cite{colleoni2014echo}, social bots can be used to interfere in the political dialogue as a biased attack vector for opinion manipulation.  For instance, some works showed evidence of the prevalence of bots in the 2016 US presidential elections \cite{boichak2018automated}, the UK-EU Brexit referendum \cite{duh2018collective}, the 2018 Italian general election \cite{stella2019influence}, and the 2019 Spanish general election \cite{pastor2020spotting}.  Social bots can be defined as automatic agents designed to mimic or impersonate humans' behavior. They are prevalent as social actors in OSN platforms, amplifying misinformation effects in several magnitudes~\cite{colleoni2014echo,lazer2018science}. Due to their artificial nature, bots have specific aims, and they do not change their opinion, neither their posture about some parties, candidates, or topics. Therefore, it is natural to wonder how the inclusion of a minimum fraction of bots could modify the behavior of tolerant and intolerant individuals and what could be the impact on a given electoral process.  How are the results of a simple model with bots compared to those obtained from ``human'' stubbornness in the voter model?

In this article, we introduce and study an extension of the voter model that incorporates bots and the tolerance level of agents.  Each agent is endowed with an opinion ($A,B$) and a tolerance ($+,-$) that is updated according to the voter dynamics.  The opinion and tolerance processes are coupled to each other and take place on two different networks, forming a multiplex network topology.  The dynamic on the opinion layer is affected by that of the tolerance layer by a mechanism that makes intolerant agents more resilient to switch opinion.  This framework also allows the introduction of bots, modeled as agents that try to change other agents' opinions but are not influenced by them.  Thus, these bots can be seen as stubborn agents that try to model the presence of opinion makers or the use of a false profile by political actors on a social network to influence electoral results.

We need to mention that some previous related works have also implemented voter-like dynamics on multiplex networks \cite{Velasquez-2017,DaSilva-2019,Velasquez-2020,Wang-2014,Granell-2014}.  However, the models in these works explore how the propagation of an opinion, rumor, or information affects the spreading of a disease in a population.  Therefore, they couple the voter dynamics in one layer with that of the $SI$, $SIS$ or $SIR$ dynamics in the other layer ($S$, $I$ and $R$ stands for susceptible, infected and recovered individuals), unlike in our model where both layer support a voter dynamics.

The rest of the article is organized as follows.  In Section~\ref{sec:model} we define the model and its dynamics on a bi-layer network.  In Section~\ref{sec:MF-results} we develop a mean-field approach to study the version of the model without bots.  We perform a stability analysis of the steady states and estimate the consensus times.  Section~\ref{sec:bots} is dedicated to the study of the model with bots.  Results from Monte Carlo simulations are presented in Section~\ref{sec:results}.  Finally, in Section~\ref{sec:conclsuions} we summarize and give the conclusions.

\section{\label{sec:model}Multilayer Voter model}

\begin{figure}[]
\centering
\resizebox{0.85\textwidth}{!}{%
\includegraphics{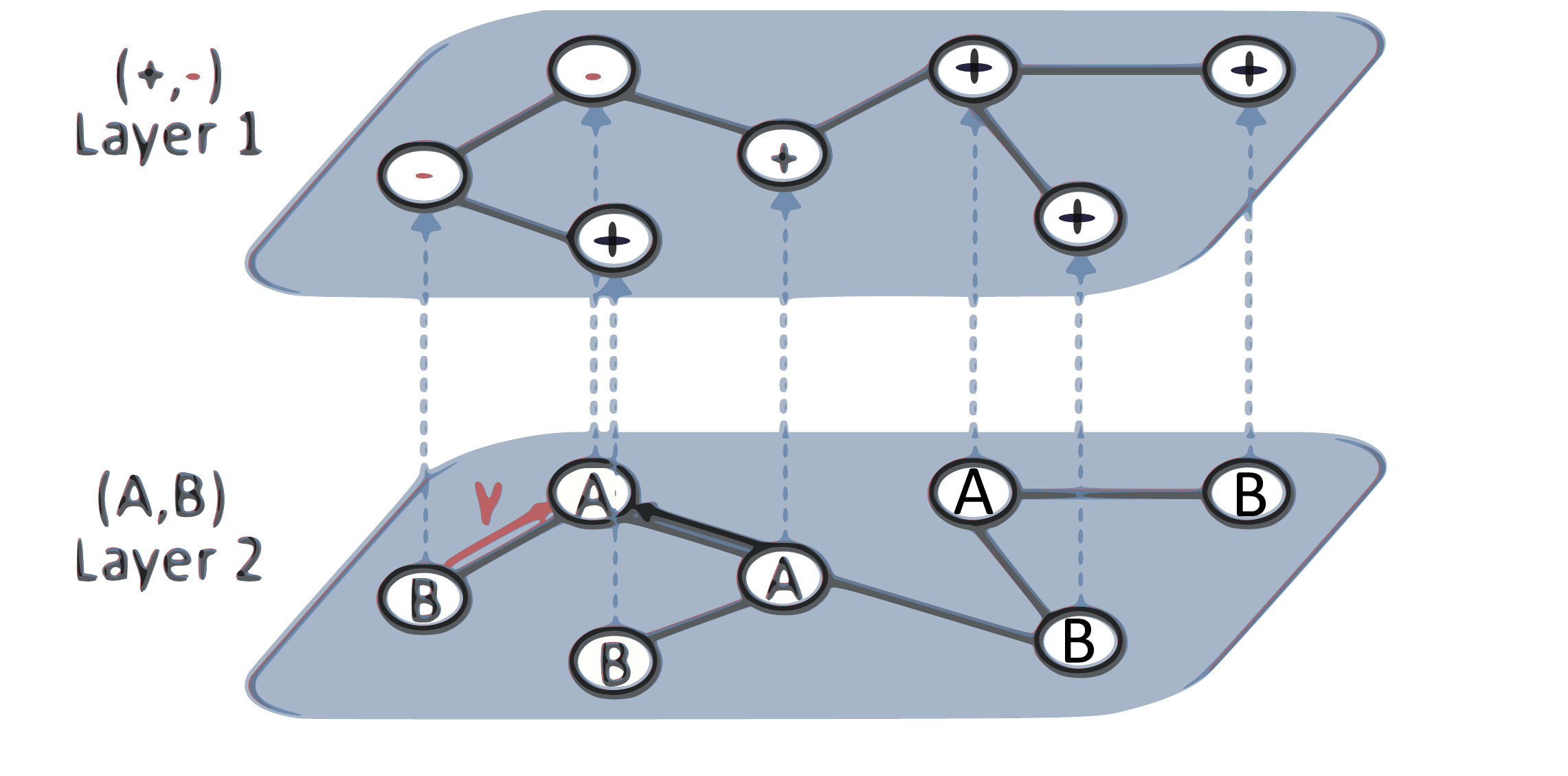}
}
\caption{\small Schematic representation of the bi-layer voter model.  The network of interactions between agents' opinions ($A$ and $B$) is represented in layer $2$, while layer $1$ represents the interactions between tolerance levels ($+$ and $-$) of agents. Opinion and tolerance states are updated according to the voter dynamics, i.e., by copying the state of a random neighbor in the corresponding layer.  A tolerant agent (tolerance $+$) copies a neighbor's opinion with probability $1$, while an intolerant agent (tolerance $-$) adopts the opinion of a neighbor with a smaller probability $\gamma \le 1$, and becomes tolerant.}
\label{fig:initial}
\end{figure}

We consider a population of interacting agents in which each agent can adopt one of two possible opinions \gO{O} = $A$ or $B$.  Besides, agents are endowed with a tolerance value \gO{T} = $+$ or $-$ that indicates the willingness of an agent to change its opinion, where the positive posture ($+$) means that the agent is more tolerant and open to switching between both opinions, and the negative posture ($-$) indicates that the agent is more radical or convinced about its own opinion, and thus less likely to change.  The system of agents and their interactions are represented by a multiplex network composed of two layers of networks with an equal number of nodes (see Figure~\ref{fig:initial}), where nodes in layers $1$ (tolerance layer $\pm$) and $2$ (opinion layer $AB$) describe the tolerance and opinions of agents, respectively.  The multiplex topology means that each node in the $\pm$--layer is connected to a node in $AB$--layer by an inter-layer link (dashed vertical arrow), representing an agent's opinion and its tolerance, but the configuration of links within each network layer (intra-layer connections) could be different.  Besides, we consider that the networks have no degree correlations, i.e., nodes are randomly connected.  

To simplify notation, we denote by $\tiny \braces{\gO{T}}{\gO{O}}$ the state of a node in the bi-layer system, and thus there are four possible node states: 
\begin{equation}
\braces{\gO{T}}{\gO{O}} =  \left\{ \braces{+}{A}, \braces{-}{A}, \braces{+}{B}, \braces{-}{B} \right\}.
\label{eq:hetSp}
\end{equation} 
In a single time step $\Delta t=1/N$ of the dynamics, a node $i$ with state $\tiny \braces{\gO{T}_i}{\gO{O}_i}$ is chosen at random, and its tolerance $\gO{T}_i$ and opinion $\gO{O}_i$ are updated according to the voter dynamics.  That is, a random neighbor $j$ with state $\tiny \braces{\gO{T}_j}{\gO{O}_j}$ is chosen from the $\pm$--layer, and a random neighbor $k$ with state $\tiny \braces{\gO{T}_k}{\gO{O}_k}$ is chosen from the $AB$--layer.  Then, node $i$ copies the tolerance of node $j$ in layer $\pm$ ($\gO{T}_i \rightarrow \gO{T}_i=\gO{T}_j$): 
\begin{equation}
\braces{\gO{T}_i}{\gO{O}_i}  \ \braces{\gO{T}_j}{\gO{O}_j} \stackrel{1}{\longrightarrow} \braces{\gO{T}_j}{\gO{O}_i} \ \braces{\gO{T}_j}{\gO{O}_j}.
\label{eq:r1}
\end{equation}
Also, node $i$ copies the opinion of node $k$ in layer $AB$ ($\gO{O}_i \rightarrow \gO{O}_i=\gO{O}_k$) with probability $1$ if its tolerance is $\gO{T}_i=+$:
\begin{equation}
\braces{+}{\gO{O}_i}  \ \braces{\gO{T}_k}{\gO{O}_k} \stackrel{1}{\longrightarrow} \braces{+}{\gO{O}_k} \ \braces{\gO{T}_k}{\gO{O}_k},
\label{eq:r3}
\end{equation}
and with probability $\gamma$ if $\gO{T}_i=-$:
\begin{eqnarray}
& \braces{-}{\gO{O}_i} & \ \braces{\gO{T}_k}{\gO{O}_k}  \stackrel{\gamma}{\longrightarrow}  \braces{+}{\gO{O}_k}  \ \braces{\gO{T}_k}{\gO{O}_k} ~~~\mbox{when $\gO{O}_i \neq \gO{O}_k$} ~~~ \mbox{and} \\
& \braces{-}{\gO{O}_i} & \ \braces{\gO{T}_k}{\gO{O}_k}  \stackrel{1}{\longrightarrow}  \braces{-}{\gO{O}_k}  \ \braces{\gO{T}_k}{\gO{O}_k} ~~~\mbox{when $\gO{O}_i = \gO{O}_k$}
\label{eq:r4}
\end{eqnarray}
Table~\ref{table-no-bots} in \ref{sec:transitions} shows explicitly all possible transitions when a pair of nodes interact. 

In other words, agents adopt the tolerance of a random neighbor in the tolerance $\pm$--layer, following a known mechanism called social influence by which a tolerant individual tends to become intolerant or radical when most of their acquaintances are intolerant, and vice-versa.  In the opinion $AB$--layer, each agent copies a random neighbor's opinion with probability $1$ if it is tolerant, but with probability $\gamma$ if it is intolerant.  This tries to capture the fact that intolerant or radical individuals are less likely to change opinion than tolerant or moderate individuals, which is modeled by assuming that intolerant agents change their minds with a reduced probability $\gamma \le 1$.  Additionally, if an intolerant agent does change opinion, we assume that it also becomes tolerant, as it is expected that a radical individual that changes its mind is prone to become more tolerant or, similarly, it is rarely expected that radical individuals suddenly adopt a radical position of the opposite view.  As we can see, the dynamics of the two layers affect each other.  On the one hand, the $\pm$--layer influences the dynamics on the $AB$--layer by reducing the rate at which intolerant agents switch opinion.  On the other hand, the $AB$--layer influences the tolerance states in the $\pm$--layer by turning intolerant agents to tolerant when they change opinion. 

\section{\label{sec:MF-results}Mean-field approach}

The state of the system at the macroscopic level is well characterized by the global densities of nodes in each of the four tolerance--opinion states, $\sigma_{\tiny \gO{O}}^{\tiny \gO{T}} = \sigma_A^+, \sigma_A^-, \sigma_B^+$ and $\sigma_B^-$.  Given that the number of nodes is conserved in each layer, the conditions  
$\sigma_A^+(t) + \sigma_A^-(t) + \sigma_B^+(t) + \sigma_B^-(t) = 1$, $\sigma_A(t) + \sigma_B(t)=1$ and  $\sigma^+(t) + \sigma^-(t) = 1$ must be fulfilled for all time $t \ge 0$, where $\sigma_{\tiny \gO{O}} = \sigma_{\tiny \gO{O}}^+ + \sigma_{\tiny \gO{O}}^-$ and $\sigma^{\tiny \gO{T}} = \sigma_A^{\tiny \gO{T}} + \sigma_B^{\tiny \gO{T}}$ are the density of agents with opinion $\gO{O}$ and tolerance $\gO{T}$, respectively.  Within a mean-field (MF) approach, the time evolution of the densities is given by the following set of rate equations:
\begin{subequations}
  \begin{alignat}{4}
    \label{eq:A+}
	\frac{d\sigma_{A}^+}{dt} = & \sigma_{A}^-\sigma^+ + \gamma\sigma_{B}^-\sigma_{A} + \sigma_{B}^+\sigma_{A} - \sigma_{A}^+\sigma^- - \sigma_{A}^+\sigma_{B},  \\ 
	\frac{d\sigma_{A}^-}{dt} = & \sigma_{A}^+\sigma^- - \sigma_{A}^-\sigma^+ - \gamma\sigma_{A}^- \sigma_{B}, \\ 
	\frac{d\sigma_{B}^+}{dt} = & \sigma_{B}^-\sigma^+ + \gamma\sigma_{A}^-\sigma_{B} + \sigma_{A}^+\sigma_{B} - \sigma_{B}^+\sigma^- - \sigma_{B}^+\sigma_{A}, \\ 
	\frac{d\sigma_{B}^-}{dt} = & \sigma_{B}^+\sigma^- - \sigma_{B}^-\sigma^+ - \gamma\sigma_{B}^-\sigma_{A}.
  \end{alignat}
\label{eq:voterModel}
\end{subequations}
This approach neglects state correlations between neighboring nodes in the networks. It thus should work reasonably well for random networks with homogeneous degree distributions and without degree correlations, such as the Erd\"{o}s-R\'enyi networks.  The gain and loss terms in Eqs.~(\ref{eq:voterModel}) correspond to the different transitions between node states.  For instance the gain term $\sigma_{A}^- \sigma^+$ in Eq.~(\ref{eq:A+}) corresponds to the transition of a node to state $\tiny \braces{-}{A}$ to state $\tiny \braces{+}{A}$ in a time step, when its tolerance switches from $-$ to $+$: a $\tiny \braces{-}{A}$--node $i$ is chosen with probability $\sigma_{A}^-$ and copies the tolerance of a random neighbor $j$, which has tolerance $\gO{T}_j=+$ with probability $\sigma^+$.  Within an MF approximation, we are assuming here that the fraction of neighbors of node $i$ with tolerance $\gO{T}=+$ is approximately equal to $\sigma^+$.  

Expanding the expressions for $\sigma^+, \sigma^-, \sigma_A$ and $\sigma_B$ in Eq.~(\ref{eq:voterModel}) in terms of the four densities $\sigma_{A}^+, \sigma_{A}^-, \sigma_{B}^+$ and $\sigma_{A}^-$ we obtain, after rearranging terms, the following closed system of rate equations:
\begin{subequations}
  \begin{alignat}{4}
    \label{A+2}
	\frac{d\sigma_{A}^+}{dt} = & 
	2\sigma_{A}^- \sigma_B^+ + \gamma  \sigma_{B}^-(\sigma_A^+ + \sigma_A^-)  -
	2\sigma_{A}^+ \sigma_B^-,\\ 
    \label{A-2}	
	\frac{d\sigma_{A}^-}{dt} = &
	\sigma_{A}^+\sigma_B^- -
	\sigma_{A}^-\sigma_B^+ -
	\gamma\sigma_{A}^- (\sigma_B^+ + \sigma_B^-),\\
    \label{B+2}	
	\frac{d\sigma_{B}^+}{dt} = &
	2\sigma_{B}^-\sigma_A^+ +
	\gamma \sigma_{A}^- (\sigma_B^+ + \sigma_B^-) -
	2\sigma_{B}^+\sigma_A^-,\\
    \label{B-2}	
	\frac{d\sigma_{B}^-}{dt} = &
	\sigma_{B}^+\sigma_A^- -
	\sigma_{B}^-\sigma_A^+ -
	\gamma\sigma_{B}^-(\sigma_A^+ + \sigma_A^-).
  \end{alignat}
\label{eq:voterModel2}
\end{subequations}	
To study the behavior of the multilayer system, we numerically integrated Eqs.~(\ref{eq:voterModel2}) subject to the symmetric initial condition in opinion $\sigma_A(0) = \sigma_B(0) = 0.5$ and tolerance $\sigma^+(0) = \sigma^-(0) = 0.5$, and for six different values of $\sigma_B^-(0)=0.25, 0.3, 0.35, 0.4, 0.45$ and $0.5$, so that the other three node densities are $\sigma_A^+(0)=0.5-\sigma_B^+(0)=\sigma_B^-(0)$ and  $\sigma_A^-(0)=\sigma_B^+(0)=0.5-\sigma_B^-(0)$.  In order to explore how radical agents of a given opinion affect the final outcome of the model, we are considering an initial state that favors intolerant agents with opinion $B$ ($\sigma_B^-(0) \ge 0.25$), compared to the perfectly symmetric condition $\sigma_A^+(0)=\sigma_A^-(0)=\sigma_B^+(0)=\sigma_B^-(0)=0.25$.

\subsection{\label{steady}Steady states}

 The system of Eqs.~(\ref{eq:voterModel2}) has four trivial fixed points $(\sigma_A^+,\sigma_A^-,\sigma_B^+,\sigma_B^-) = $(1,0,0,0)$, $ $(0,1,0,0)$, $(0,0,1,0)$ and $(0,0,0,1)$ corresponding to a consensus in states $A^+$, $A^-$, $B^+$ and $B^-$, respectively.  These are absorbing (inactive) states where there are no more possible updates, as all agents have the same opinion and tolerance.  Besides, Eqs.~(\ref{eq:voterModel2}) have infinitely many non-trivial fixed points $\vec{\sigma}^* = (1-\sigma_B^*,0,\sigma_B^*,0)$ that correspond to a consensus of tolerant agents ($\sigma^+=1, \sigma^-=0$), where $\sigma_B^* = \sigma_B(t=\infty)=\sigma_B^+(t=\infty)$ ($\sigma_B^* \in [0,1]$) is the stationary density of agents with opinion $B$. As there are only agents with $+$ tolerance at the steady state, we have  $\sigma_A(t=\infty)=\sigma_A^+(t=\infty)=1-\sigma_B^*$.  This can be considered as a steady state of coexistence between $A$ and $B$ tolerant agents, with constant densities over time.  This happens because the system is reduced to a simple $2$-state symmetric voter model where the fraction of voters that make a transition from state $A^+$ to state $B^+$ per unit time, $\sigma_A^+ \sigma_B^+$, is equal to the fraction of voters making the reverse transition (from $B^+$ to $A^+$), thus the net flow is zero and the densities are conserved.
 
 \begin{figure}[]
\centering

\subfigure[]{\includegraphics[scale=0.22]{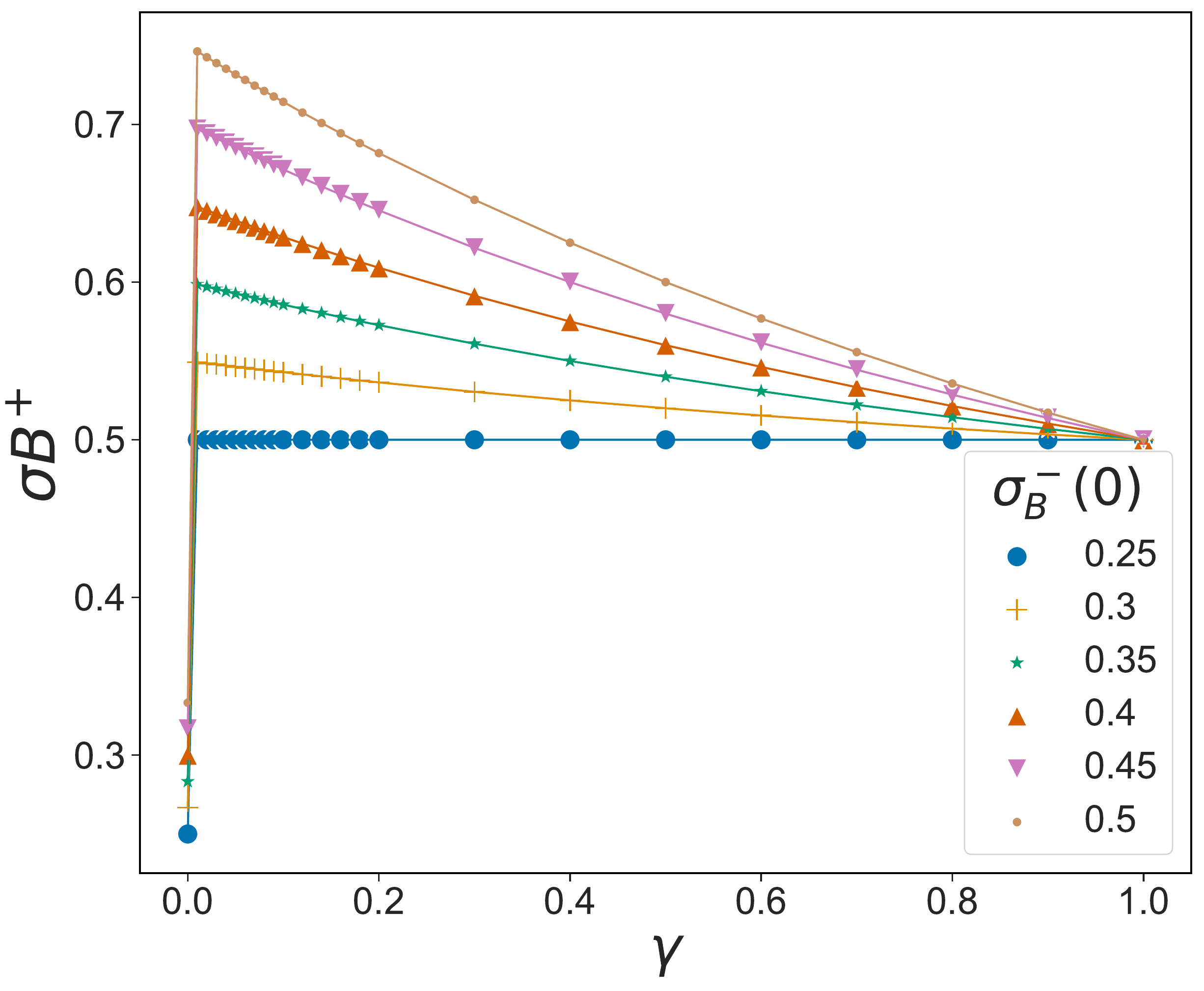}}\hspace{0.5cm}
\subfigure[]{\includegraphics[scale=0.22]{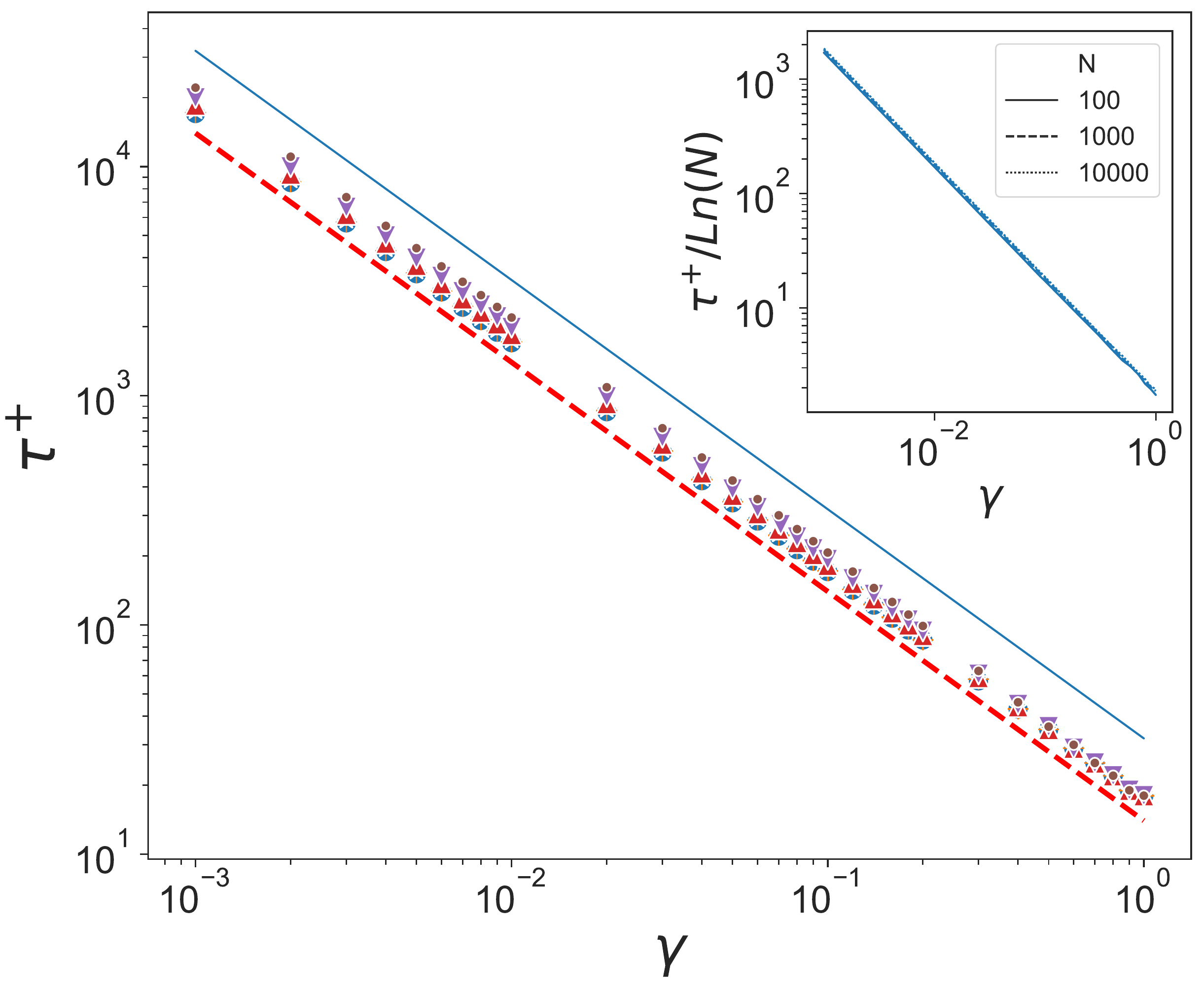}}
\caption{\small Effects of the initial conditions, tolerance level $\gamma$ and the initial density of bias radical individuals $\sigma_{B}^-$, in the outcome of the bi-layer voter model without bots: (a) the final density of tolerant agents with opinion B  ($\sigma_{B}^+$); (b) the consensus time of tolerant agents with respect of $\gamma$. The red dotted line below is the $1/\gamma$ curve.}
\label{fig:impact}
\end{figure}
 
 We have checked that the density of tolerant agents with opinion $B$ at the stationary state $\sigma_B^*$ depends on the initial condition, controlled by the initial density of opinion $B$ intolerant agents $\sigma_B^-(0)$.  This can be seen in Figure~\ref{fig:impact}(a), where we plot $\sigma_B^*$ vs the likelihood $\gamma$ of intolerant agents to change opinion, for various values of $\sigma_B^-(0)$.  We observe that, for a fixed value of $\gamma$, $\sigma_B^*$ increases with $\sigma_B^-(0)$, meaning that a larger initial number of $B$--agents leads to a larger final number of $B$--agents.  We also see a more intriguing effect, that $\sigma_B^*$ increases as $\gamma$ decreases.  We can obtain an insight into these results from a closer inspection of Eqs.~(\ref{eq:voterModel2}).  Adding Eqs.~(\ref{B+2}) and (\ref{B-2}) we obtain that the density of opinion $B$ agents evolves according to
 \begin{equation}
     \frac{d \sigma_B}{dt} = \left( 1-\gamma) (\sigma_A^+ \sigma_B^- - \sigma_A^- \sigma_B^+ \right),
     \label{dsBdt}
 \end{equation}
 while adding Eqs.~(\ref{A-2}) and (\ref{B-2}) leads to the following evolution of the density of intolerant ($-$) agents:
\begin{equation}
     \frac{d \sigma^-}{dt} = - \gamma \left( \sigma_A^+ \sigma_B^- + 2 \sigma_A^- \sigma_B^- + \sigma_A^- \sigma_B^+ \right).
     \label{ds-dt}    
\end{equation} 
Given that all four initial densities can be written in terms of $\sigma_B^-(0)$, we arrive from Eq.~(\ref{dsBdt}) that at $t=0$ is 
\begin{equation}
    \frac{d \sigma_B(0)}{dt} = (1-\gamma) \left[ \sigma_B^-(0) - 0.25 \right],
    \label{dsdt0}
\end{equation}
which is larger than zero for all initial conditions $\sigma_B^-(0) > 0.25$ of Figure~\ref{fig:impact}(a).  Therefore, it is expected that, for any $\gamma>0$, $\sigma_B$ increases from $0.5$ at $t=0$ to a stationary value $\sigma_B^*$ larger than $0.5$ as $t \to \infty$, explaining why all curves of Figure~\ref{fig:impact}(a) are above $0.5$, except the initially symmetric case $\sigma_B^-(0)=0.25$ for which the densities are conserved.  Another exception is the $\gamma=1$ case, where opinion densities are conserved [see Eq.~(\ref{dsBdt})], and so $\sigma_B(t)=\sigma_B(0)=0.5$ and $\sigma_A(t)=\sigma_A(0)=0.5$ for all $t \ge 0$.

As we described above, the initial asymmetric state that favors $B^-$ agents leads to a stationary state with a majority of $B$--agents ($\sigma_B^*>0.5$).  This behavior is more pronounced as $\gamma$ decreases [Eq.~(\ref{dsdt0})], and it seems to be the reason why $\sigma_B^*$ increases as $\gamma$ approaches zero, as we see in Figure~\ref{fig:impact}(a), showing a maximum (peak) in the $\gamma \to 0$ limit.

The $\gamma=0$ case is special because the tolerance densities are conserved [see Eq.~(\ref{ds-dt})], and so $\sigma^+(t)=\sigma^+(0)=0.5$ and $\sigma^-(t)=\sigma^-(0)=0.5$ for all $t \ge 0$.  As a consequence, $\sigma_A^+=0.5-\sigma_B^+$ and $\sigma_A^-=0.5-\sigma_B^-$ for $t \ge 0$.  Replacing these expressions for $\sigma_A^+$ and $\sigma_A^-$ in Eq.~(\ref{dsBdt}) we obtain
\begin{equation}
    \frac{d \sigma_B}{dt} = 0.5 \left( \sigma_B^- - \sigma_B^+ \right), 
    \label{dsBdt0}
\end{equation}
and thus 
\begin{equation}
    \frac{d \sigma_B(0)}{dt} = 0.5 \left[ 2 \sigma_B^-(0) - 0.5 \right]     
\end{equation}
at $t=0$.  Then, for $\sigma_B^-(0) \ge 0.25$ we expect that $\sigma_B$ increases from $0.5$ at $t=0$ and reaches a value $\sigma_B^* \ge 0.5$.  Finally, given that $\sigma_B^+=\sigma_B^-$ at the stationary state [Eq.~(\ref{dsBdt0})], we have that $2\sigma_B^+=\sigma_B^* \ge 0.5$ and thus $\sigma_B^+ \ge 0.25$, as we can check in Figure~\ref{fig:impact}(a) for $\gamma=0$.

For $\gamma>0$, the right-hand-side of Eq.~(\ref{ds-dt}) is always negative, thus $\sigma^-$ decreases and eventually approaches zero in the $t \to \infty$ limit, corresponding to a consensus in the tolerant state ($+$) at the steady-state ($\sigma^-=0$, $\sigma^+=1$) as we mentioned before.  A magnitude of interest is the time to reach the tolerant consensus $\tau^+$.  Given that the rate Eqs.~(\ref{eq:voterModel2}) describe an infinitely large system where finite-size fluctuations are neglected, we estimated $\tau^+$ as the time for which the density of tolerant ($+$) agents becomes larger than $1-1/N$, that is, when there is less than one agent with state $-$.  Results are shown in Figure~\ref{fig:impact}(b) where we plot $\tau^+$ as a function of $\gamma$ for different initial conditions.  We can see that $\tau^+$ diverges as $\tau^+ \sim 1/\gamma$ when $\gamma$ approaches zero.  The intuition behind this result is that for $\gamma \ll 1$ the consensus time is determined by the slowest time scale of the system, associated with the transition of all intolerant agents $-$ to the tolerant state $+$ at rate $\gamma$, which takes a time of order $1/\gamma$.  
We also see that $\tau^+$ is not strongly affected by the initial state that favors $B^-$ agents ($\sigma_B^-(0)>0.25$).

In summary, these results show that the system eventually reaches a tolerant consensus in the long run. Still, the convergence could be extremely slow when radical agents are unlikely to change their opinion, and that it becomes infinitely large (there is never consensus) for the extreme case of stubborn or intolerant agents ($\gamma=0$).

\subsection{\label{stability} Stability analysis and consensus times} 

A better estimation of the tolerant consensus time  $\tau^+$ can be obtained from a linear stability analysis of the tolerant fixed point $\vec{\sigma}^* = (1-\sigma_B^*,0,\sigma_B^*,0)$.  For that, we consider small perturbations $\epsilon_i$ ($i=1,2,3,4$) of the components of $\vec{\sigma}^*$ and write $\sigma_A^+= 1-\sigma_B^* + \epsilon_1$, $\sigma_A^-=\epsilon_2$, $\sigma_B^+=\sigma_B^*+\epsilon_3$ and $\sigma_B^-=\epsilon_4$, where $\sum_{i=0}^4 \epsilon_i=0$.  Inserting these expressions for the densities into Eqs.~(\ref{eq:voterModel2}) and neglecting terms of order $2$ we obtain, to first order in $\epsilon_i$, the following system of linear equations in matrix representation: 
\begin{eqnarray*}
  \frac{d \vec{\epsilon}}{dt} = {\bf A} \, \vec{\epsilon},
\end{eqnarray*}
where  
\begin{eqnarray*}
  {\bf A} \equiv \left( {\begin{array}{cccc}
   0 & 2 \sigma_B^* & 0 & (1-\sigma_B^*)(\gamma-2)) \\
   0 & -\sigma_B^* (1+\gamma) & 0 & 1-\sigma_B^* \\
   0 & \sigma_B^*(\gamma-2) & 0 & 2(1-\sigma_B^*) \\
   0 & \sigma_B^* & 0 & - (1-\sigma_B^*)(1+\gamma)
  \end{array} } \right),
\end{eqnarray*}
and $\vec{\epsilon} \equiv (\epsilon_1,\epsilon_2,\epsilon_3,\epsilon_4)$.  Matrix ${\bf A}$ has two negative eigenvalues
\begin{equation}
    \lambda_{\pm} = \frac{-1 - \gamma \pm \sqrt{(1+\gamma)^2 - 4\sigma_B^*(1-\sigma_B^*)\gamma(2+\gamma)}}{2}
\end{equation}
associated with a perturbation in the total densities of $+$ and $-$ agents $1.0$ and $0$, respectively, but that keeps the densities of $A$ and $B$--agents $1-\sigma_B^*$ and $\sigma_B^*$, respectively, unchanged.  This means that the tolerance consensus state is stable.  The other two eigenvalues are zero.  One corresponds to the conservation of the total density of agents $1.0$, and the other describes the instability of $\vec{\sigma}^*$ after a perturbation that changes the densities of $A$ and $B$--agents.  Then, the perturbations evolve according to $\epsilon_i = a_i + b_i \, e^{\lambda_+ t} + c_i \, e^{\lambda_- t}$, where $a_i, b_i$ and $c_i$ are constants given by the initial condition, and thus the density of tolerant agents $\sigma^+=\sigma_A^+ + \sigma_B^+$ evolves after a smaller perturbation as
\begin{equation}
    \sigma^+(t) \simeq 1 + \epsilon_1(t) + \epsilon_3(t) = 1 + (a_1+a_3) + (b_1+b_3) e^{\lambda_+ t} + (c_1+c_3) e^{\lambda_- t}.
    \label{sigma-t}
\end{equation}
As we know that $\sigma^+$ approaches $1$ as $t \to \infty$ [see Eq.~(\ref{ds-dt})] and that $\lambda_-< \lambda_+ < 0$ for $\gamma>0$, the coefficient corresponding to the $0$ eigenvalue $a_1+a_3$ must be zero.  Besides, at long times only the term corresponding to the largest eigenvalue $\lambda_+$ survives (smallest absolute value), and thus Eq.~(\ref{sigma-t}) becomes
\begin{equation}
    \sigma^+(t) \simeq 1 + (b_1+b_3) e^{\lambda_+ t}.
    \label{sigma-t-2}
\end{equation}
The time to reach consensus can be estimated from Eq.~(\ref{sigma-t-2}) as the time $\tau^+$ for which the density of tolerant $+$ agents reaches the value $1-1/N$, that is, $\sigma^+(\tau) = 1 + (b_1+b_3) e^{\lambda_+ \tau} = 1-1/N$, from where we arrive at the approximate expression
\begin{equation}
    \tau^+ \simeq \frac{\ln \left[-(b_1+b_3) N \right]}{-\lambda^+}.
    \label{tau+}
\end{equation}
We notice that, as $b_1+b_3<0$ [Eq.~(\ref{sigma-t-2})] and $\lambda_+<0$, expression Eq.~(\ref{tau+}) gives a physical time $\tau^+ > 0$.  In Figure~\ref{fig:impact}(b) we see that the approximate expression Eq.~(\ref{tau+}) (solid lines) captures quite well the behavior of $\tau^+$ with $\gamma$ obtained from the integration of Eqs.~(\ref{eq:voterModel2}) (symbols).

\section{\label{sec:bots} Inclusion of Bots}

We now include in the model a fraction of Bots $\sigma_{\mathbb{B}}^-$ that remains constant over time.  Bots are artificial entities that diffuse opinions related to a specific position.  Due to their artificial nature, bots do not change opinion neither the posture.  In this section, we analyze the effects of including bots that have a fixed opinion $B$, and so they can be considered as extremist intolerant agents in the state $B^-$.  The total density of agents is now decomposed in five terms,

\begin{equation}
\begin{array}{lc}
\sigma_{A}^+ + \sigma_{A}^- + \sigma_{B}^+ + \sigma_{B}^- + \sigma_{\mathbb{B}}^- = & 1, \\
\end{array}
\label{eq:botrules}
\end{equation}
where $\sigma_{\mathbb{B}}^-(t) = \sigma_{\mathbb{B}}^-(0)$ for all $t \ge 0$.  We also consider the same initial conditions as that without bots, determined by $\sigma_B^-(0)$, i.e.,
{$\sigma_A^+(0)=\sigma_B^-(0)$ and  $\sigma_A^-(0)=\sigma_B^+(0)=0.5-\sigma_B^-(0)$, leading to $\sigma^+(0) = 0.5$ and $\sigma_A(0) = 0.5$.}
The rates equations for the evolution of the densities can be derived following the same procedure as that for the model with no bots at the beginning of Section~\ref{sec:MF-results}, considering an extra compartment $\mathbb{B}^-$ that behaves as an intolerant state $B^-$, but with the important distinction that transitions from state $\mathbb{B}^-$ to states $B^+$ and $A^+$ are not allowed (see table~\ref{table-bots} in \ref{sec:transitions} for a detailed description of all possible transitions).  We make clear that agents ``see'' a bot as another $B^-$ agent. Still, they make transitions only between the four states $A^+, A^-, B^+$ and $B^+$ (never to the bots' state $\mathbb{B}^-$), so that the total density of agents $1-\sigma_{\mathbb{B}}^-$ as well as the density of bots $\sigma_{\mathbb{B}}^-$ are conserved quantities.  The resulting set of MF equations reads
 
\begin{subequations}
  \begin{alignat}{4}
    \label{A+b}
	\frac{d\sigma_{A}^+}{dt} = & 
	2\sigma_{A}^- \sigma_B^+ + \gamma \sigma_B^-(\sigma_A^+ + \sigma_A^-) -
	2 \sigma_{A}^+ (\sigma_B^- + \sigma_{\mathbb{B}}^-),
	\\
	\label{A-b}
    \frac{d\sigma_{A}^-}{dt} = &
	\sigma_{A}^+ (\sigma_B^- + \sigma_{\mathbb{B}}^-) -
	\sigma_{A}^-\sigma_B^+ -
	\gamma\sigma_{A}^- (\sigma_B^+ + \sigma_B^- + \sigma_{\mathbb{B}}^-),
	\\ 
	\label{B+b}
	\frac{d\sigma_{B}^+}{dt} = &
	\sigma_A^+ (2 \sigma_B^- + \sigma_{\mathbb{B}}^-) +
	\gamma\sigma_{A}^-(\sigma_B^+ + \sigma_B^- + \sigma_{\mathbb{B}}^-) -
	\sigma_{B}^+ (2 \sigma_A^- + \sigma_{\mathbb{B}}^- ),
	\\ 
	\label{B-b}
	\frac{d\sigma_{B}^-}{dt} = &
	\sigma_{B}^+ (\sigma_A^- + \sigma_{\mathbb{B}}^-) -
	\sigma_B^-\sigma_A^+ -
	\gamma\sigma_B^-(\sigma_A^+ + \sigma_A^-).
  \end{alignat}
\label{eq:voterModelBots}
\end{subequations}

\subsection{\label{steady-nobots}Steady states}

\begin{figure}[]
\centering
\subfigure[]{\includegraphics[scale=0.22]{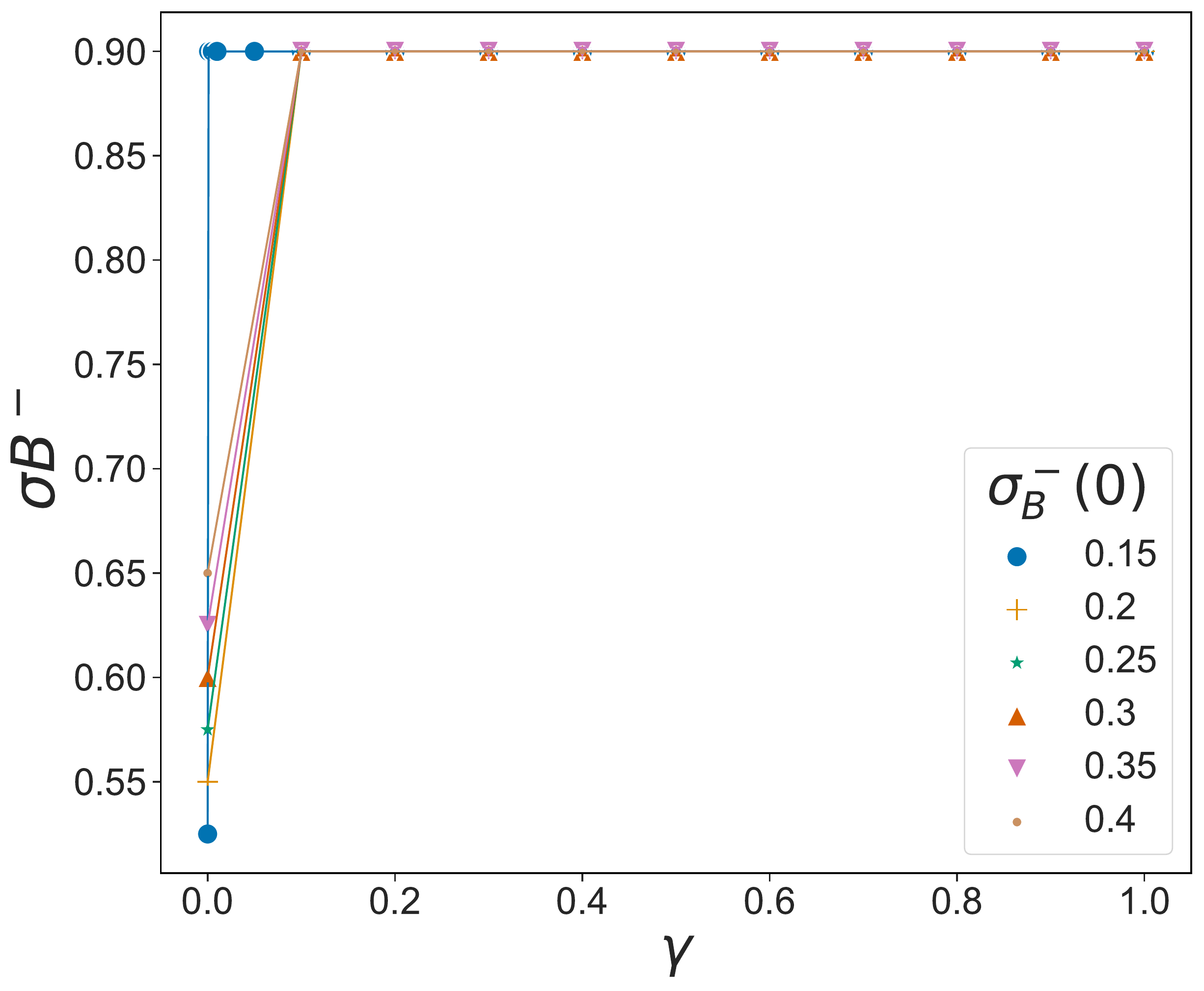}}
\subfigure[]{\includegraphics[scale=0.22]{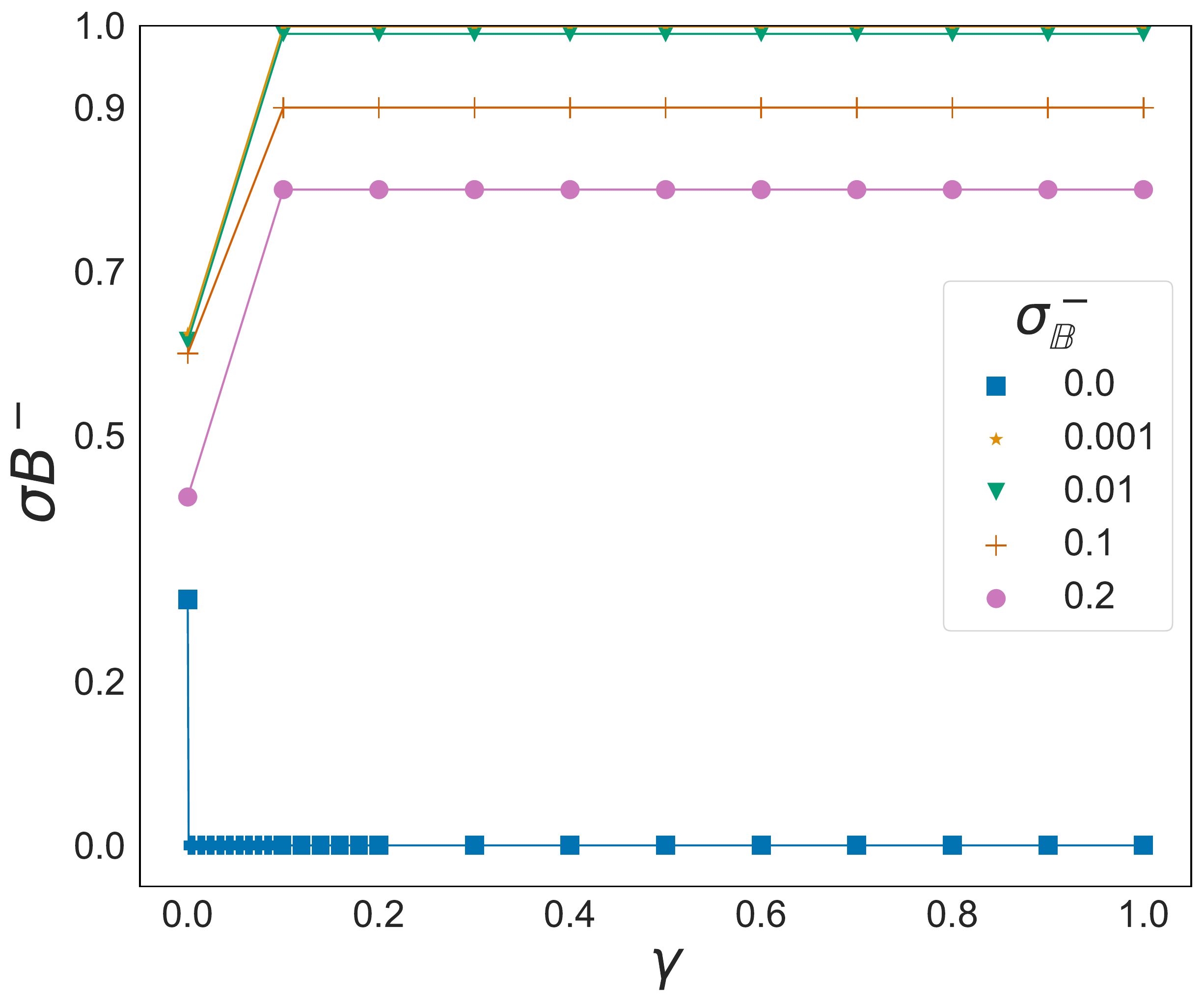}}
\subfigure[]{\includegraphics[scale=0.22]{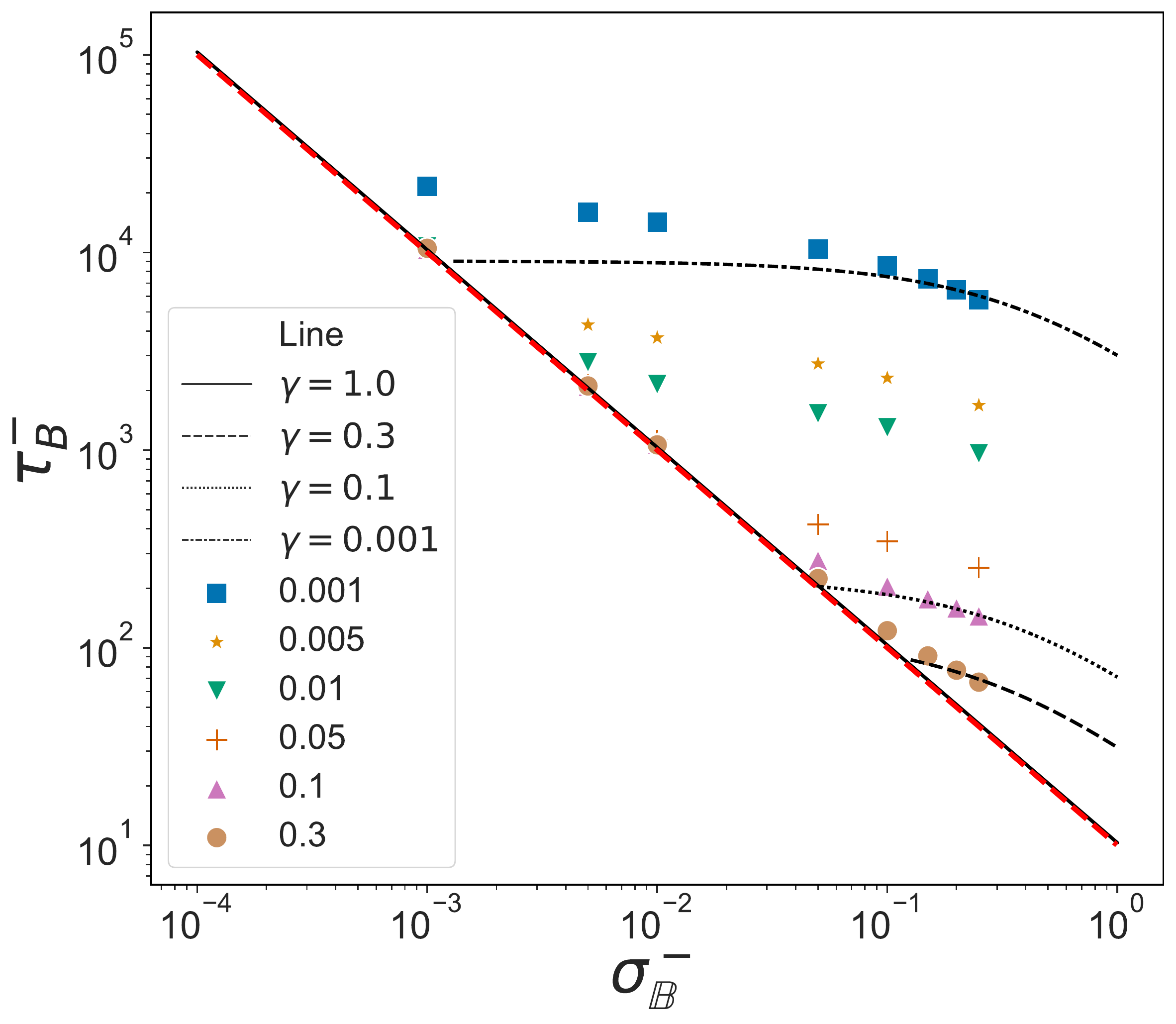}}
\subfigure[]{\includegraphics[scale=0.22]{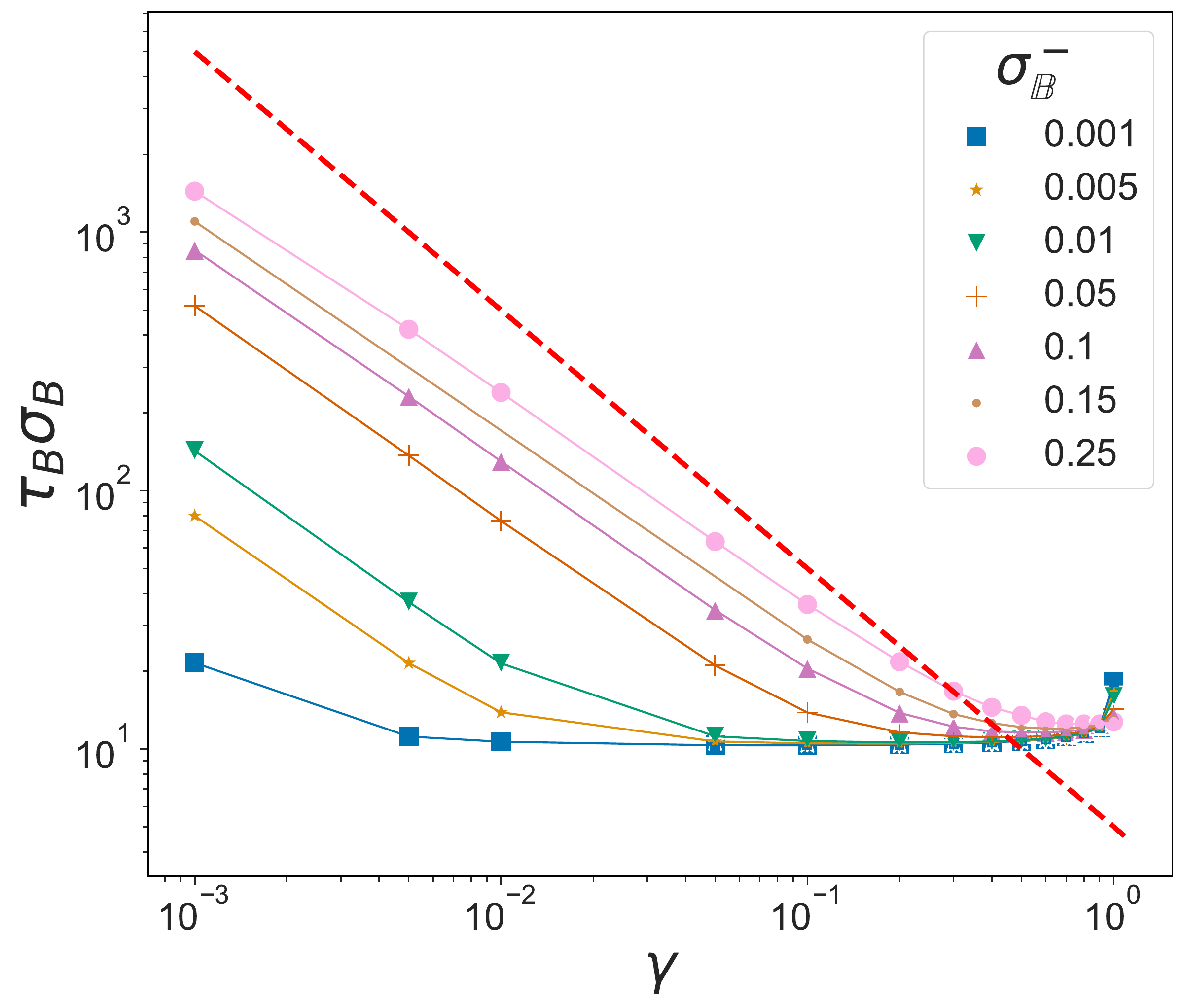}}
\caption{\small Effects of the initial conditions in the outcome of the bi-layer voter model with the inclusion of bots. (a) The final density of intolerant agents B when the density of bots is $\sigma_{\mathbb{B}}^- = 0.1$ and different bias densities of intolerant B. (b) The effect of the density of bots in the final density of intolerant B agents, with no initial intolerance bias, i.e., $\sigma_{B}^+ = 0.25$. (c) The numerical and approximated consensus time to reach the final density of intolerant B. The dotted line represents $ 1 / \sigma_{\mathbb{B}}^-$. (d) The numerical consensus time to reach the final density of B agents, in terms of the density of bots $\sigma_{\mathbb{B}}^{-}$. The dotted line is $\gamma^{-1}$ curve.}
\label{fig:impactB}
\end{figure}

We integrated the rate Eqs.~(\ref{eq:voterModelBots}) for different fractions of bots $\sigma_{\mathbb{B}}^-$ and different initial conditions that favor $\sigma_B^-$, to explore how different proportions of bots, combined with tolerant agents and asymmetric initial conditions, affects the outcome of the model.  Results are shown in Figure~\ref{fig:impactB}.  In Figures~\ref{fig:impactB}(a) and (b) we observe that the stationary density of $B^-$ agents for different initial conditions and $\gamma>0$ is $\sigma_B^- = 1-\sigma_{\mathbb{B}}^-$, that is, there is always a consensus of intolerant $B$--agents, except for $\gamma=0$.  It seems that bots break the symmetry of $A$ and $B$ opinions observed in the baseline model without bots of Section~\ref{steady}, introducing a bias towards $B^-$ agents that prevents the tolerant ($+$) consensus found for the case with no bots.  Indeed, as we see in Figures~\ref{fig:impactB}(b), for the no bots case $\sigma_{\mathbb{B}}^-=0$ is $\sigma_B^-=0$, while adding a small fraction of bots is enough to remove the $+$ consensus and drive the system to the $B^-$ consensus.  For $\gamma=0$, agents that become intolerant of the $A$ opinion never escape from that state, and thus a consensus in $B^-$ is never reached.  

\subsection{\label{stability-nobots} Consensus time and stability analysis} 

In Figure~\ref{fig:impactB}(c) we plot the time to reach the $B^-$ consensus $\tau_B^-$ as a function of $\sigma_{\mathbb{B}}^-$ for various values of $\gamma$.  We see that $\tau_B^-$ decays with $\sigma_{\mathbb{B}}^-$ as $\tau_B^- \sim 1/\sigma_{\mathbb{B}}^-$ for $\sigma_{\mathbb{B}}^- \ll 1$, independent of $\gamma$ (solid line).  In Figure~\ref{fig:impactB}(d) we plot $\tau_B^-$ as a function of $\gamma$ for various values of $\sigma_{\mathbb{B}}^-$, where the $y$--axis was rescaled by $\sigma_{\mathbb{B}}^-$ to collapse the data for values of $\gamma$ close to $1.0$.  We can see that $\tau_B^- \sim C/\gamma$ for $\gamma \ll 1$, with an amplitude $C(\sigma_{\mathbb{B}}^-)$ that depends on $\sigma_{\mathbb{B}}^-$.  To gain a better understanding of these results, bellow we derive equations for the evolution of the density of $A$--agents and $+$--agents.  For that, we add Eqs.~(\ref{A+b}) and (\ref{A-b}) to obtain
\begin{equation}
    \frac{d \sigma_A}{dt} = -(1-\gamma)(\sigma_A^+ \sigma_B^- - \sigma_A^- \sigma_B^+) - \sigma_{\mathbb{B}}^- (\sigma_A^+ + \gamma \sigma_A^-),
    \label{dsAdt}
\end{equation}
and Eqs.~(\ref{A+b}) and (\ref{B+b}) to arrive at 
\begin{equation}
    \frac{d \sigma^+}{dt} = - \sigma_{\mathbb{B}}^- \sigma^+ + \gamma (\sigma_A^- \sigma_B^+ + 2\sigma_A^- \sigma_B^- + \sigma_A^+ \sigma_B^- + \sigma_{\mathbb{B}}^- \sigma_A^-). 
    \label{ds+dt}
\end{equation}
Although these equations can not be solved exactly, it proves instructive to analyze the $\gamma=1$ case, for which Eq.~(\ref{dsAdt}) adopts the simple form
\begin{equation}
    \frac{d \sigma_A}{dt} = -\sigma_{\mathbb{B}}^- \sigma_A,
\end{equation}
with solution $\sigma_A(t) = \sigma_A(0) \, e^{-\sigma_{\mathbb{B}}^- t}$.  Then, $\sigma_A$ decays exponentially fast to zero in a time that scales as $1/\sigma_{\mathbb{B}}^-$.  Once the fraction of $A$--agents is less than $1/N$ (negligible small for $N \gg 1$), the second term of Eq.~(\ref{ds+dt}) can be neglected assuming that all terms inside the brackets are of order $1/N$ (they depend on $\sigma_A^\pm$), and thus we have 
\begin{equation}
    \frac{d \sigma^+}{dt} = - \sigma_{\mathbb{B}}^- \sigma^+,
\end{equation}
from where we obtain a consensus to the $-$ state that also scales as $1/\sigma_{\mathbb{B}}^-$.  Therefore, as both the initial $B$--consensus and the subsequent $-$ consensus scale as $1/\sigma_{\mathbb{B}}^-$, we find that $\tau_B^- \sim 1/\sigma_{\mathbb{B}}^-$.  This explains the pure power law behavior of $\tau_B^-$ with $\sigma_{\mathbb{B}}^-$ for $\gamma=1$ [solid line in Figure~\ref{fig:impactB}(a)].  For $\gamma<1$ the arguments above are not valid any more, because both time scales $1/\sigma_{\mathbb{B}}^-$ and $1/\gamma$ are at play.  A more precise approach to the general case of any $\sigma_{\mathbb{B}}^-$ and $\gamma$ is given by a linear stability analysis similar to that of Section~\ref{stability} for the case without bots, as we describe below.

The only fixed point in the system of Eqs.~(\ref{eq:voterModelBots}) is $(0,0,0,1)$, corresponding to a $B^-$ consensus, as we mentioned above.  We consider a small generic perturbation of this absorbing state of the form $\sigma_A^+=\epsilon_1$, $\sigma_A^-=\epsilon_2$, $\sigma_B^+=\epsilon_3$ and $\sigma_B^-=1-\epsilon_4$, such that $\sum_{i=1}^3 \epsilon_i - \epsilon_4=0$.  The reason why we chose the $-\epsilon_4$ perturbation is to give a physical meaning to all perturbations, considering that $\epsilon_i>0$ ($i=1,..,4$), thus all densities fall in the $[0,1]$ interval, but the analysis is also valid for $\epsilon_i<0$.  Inserting these expressions for the densities into Eqs.~(\ref{eq:voterModelBots}) and expanding to first order in $|\epsilon_i| \ll 1$ we obtain $d \vec{\epsilon}/dt = {\bf A} \, \vec{\epsilon}$, where
\begin{eqnarray*}
  {\bf A} \equiv \left( {\begin{array}{cccc}
   \gamma - 2(1+\sigma_{\mathbb{B}}^-) & \gamma & 0 & 0 \\
   1+\sigma_{\mathbb{B}}^- & -\gamma(1+\sigma_{\mathbb{B}}^-) & 0 & 0 \\
   2+\sigma_{\mathbb{B}}^- & \gamma(1+\sigma_{\mathbb{B}}^-) & -\sigma_{\mathbb{B}}^- & 0 \\
   1+\gamma & \gamma & -\sigma_{\mathbb{B}}^- & 0
  \end{array} } \right).
\end{eqnarray*}
The eigenvalues of matrix {\bf A} are 
\begin{eqnarray}
    \lambda_1 &=& 0, \\
    \lambda_2 &=& - \sigma_{\mathbb{B}}^-, \\
    \label{lambda3}
    \lambda_{3,4} &=& \frac{-2-(2+\gamma)\sigma_{\mathbb{B}}^- \pm \sqrt{\left[2+(2+\gamma)\sigma_{\mathbb{B}}^- \right]^2-4\gamma(1+\sigma_{\mathbb{B}}^-)(1-\gamma+2\sigma_{\mathbb{B}}^-)}}{2}.
\end{eqnarray}
The eigenvalue $\lambda_1=0$ expresses the conservation of the total density of agents excluding bots $1-\sigma_{\mathbb{B}}^-$.  Given that $\lambda_2, \lambda_3$ and $\lambda_4$ are negative, the consensus fixed point $(0,0,0,1)$ is stable, as expected.  As we explained in Section~\ref{stability}, the consensus time is estimated by the exponential decay of the slowest mode $e^{\lambda_{\mbox{\tiny max}} t}$ ($\lambda_{\mbox{\tiny max}}<0$) to the fixed point after a perturbation, $\tau_B^- \sim -\ln N/\lambda_{\mbox{\tiny max}}$, which corresponds to the mode with the largest negative eigenvalue $\lambda_{\mbox{\tiny max}}$.  Then, given that $\lambda_4 < \lambda_3$, the consensus time is given by the largest of the two eigenvalues $\lambda_2$ and $\lambda_3$, which depends non-trivially on the relation between $\sigma_{\mathbb{B}}^-$ and $\gamma$.  That is, for a fixed value of $\gamma>0$ and decreasing $\sigma_{\mathbb{B}}^-$, we have that $\lambda_3$ approaches the value $-1+\sqrt{1-\gamma(1-\gamma)}<0$, while $\lambda_2=-\sigma_{\mathbb{B}}^-$ approaches zero from bellow.  Therefore, $\lambda_2$ becomes larger than $\lambda_3$ for $\sigma_{\mathbb{B}}^-$ small enough, and thus 
\begin{equation}
    \tau_B^- \sim \frac{\ln N}{\sigma_{\mathbb{B}}^-} ~~~ \mbox{for $\sigma_{\mathbb{B}}^- \to 0$}. 
\end{equation}
This is the behavior observed in Figure~\ref{fig:impactB}(c), where $\tau_B^-$ decays as power law of $\sigma_{\mathbb{B}}^-$ with an amplitude that is $\gamma$ independent for $\gamma \ge 0.1$ (solid line).  For $\gamma=0.001$, it seems that the values of $\sigma_{\mathbb{B}}^-$ plotted are not small enough, so we expect that $\lambda_3 > \lambda_2$, and thus $\tau_B^- \sim -\ln N/\lambda_3$.  In general, for a fixed $\gamma>0$ there is a ``crossover'' value $\hat \sigma_{\mathbb{B}}^-$ for which $\lambda_2=\lambda_3$, so that $\tau_B^-$ is determined by $\lambda_2$ for $\sigma_{\mathbb{B}}^- < \hat \sigma_{\mathbb{B}}^-$ and by $\lambda_3$ for $\sigma_{\mathbb{B}}^- > \hat \sigma_{\mathbb{B}}^-$.  This is equivalent to setting $\tau_B^-$ to the largest of the two functions $\lambda_2^{-1}$ and $\lambda_3^{-1}$ vs $\sigma_{\mathbb{B}}^-$, as plotted in Figure~\ref{fig:impactB}(c) by solid and dashed lines, respectively.  We can see that the behavior $\tau_B^- \sim 1/\sigma_{\mathbb{B}}^-$ (solid line) fits the data very well for small values of $\sigma_{\mathbb{B}}^-$, while for larger values of $\sigma_{\mathbb{B}}^-$ the behavior of $\tau_B^-$ is dominated by $\lambda_3$ (dashed lines).  Discrepancies around the crossover point $\hat \sigma_{\mathbb{B}}^-$ are due to fact that both time scales are similar close to this point, and thus $\tau_B^-$ is determined by both time scales.

A similar analysis can be done for the $\tau_B^-$ vs $\gamma$ plot [Figure~\ref{fig:impactB}(d)].  A Taylor series expansion of expression Eq.~(\ref{lambda3}) for $\lambda_3$ to first order in $\gamma$ leads to $\lambda_3 \simeq -(1/2+\sigma_{\mathbb{B}}^-) \gamma$.  Therefore, the consensus time can be approximated as 
\begin{equation}
  \tau_B^-(\gamma) \simeq \begin{cases} 
    \frac{\ln N}{(1/2+\sigma_{\mathbb{B}}^-)\gamma} ~~ \mbox{for $\gamma \lesssim \hat \gamma$}, \\ 
    \frac{\ln N}{\sigma_{\mathbb{B}}^-}~~ \mbox{$\gamma \gtrsim \hat \gamma$},
  \end{cases}
  \label{tau-gamma}
\end{equation}
where $\hat \gamma = 2\sigma_{\mathbb{B}}^-/(1+2\sigma_{\mathbb{B}}^-)$.  In Figure~\ref{fig:impactB}(d) we can see that the approximation from Eq.~(\ref{tau-gamma}) works well for $\gamma$ small (dashed line), while approximating $\tau_B^-$ as a constant of $\gamma$ for larger values of $\gamma$ is not a good estimation.  However, it seems to give the right scaling $\tau_B^- \simeq \ln N/\sigma_{\mathbb{B}}^-$ for $\gamma \lesssim 1$, as curves for different $\sigma_{\mathbb{B}}^-$ collapse into one curve when the $y$--axis is rescaled by $\sigma_{\mathbb{B}}^-$.  Indeed, at $\gamma=1$ we have $\lambda_3(\gamma=1)=-\sigma_{\mathbb{B}}^-$.

\section{\label{sec:results}Monte Carlo simulation results}

We performed extensive Monte Carlo (MC) simulations of the dynamics of the bi-layer voter model described in section~\ref{sec:model} without bots and in section~\ref{sec:bots} with bots, in order to check the results obtained with the MF approach (sections~\ref{sec:MF-results} and \ref{sec:bots}).  We run the simulations on a multiplex network composed of two networks of $N = 10^4$ nodes and mean degree $\langle k \rangle = 20$ each, which are strongly coupled to each other, i.e., every node in one network is connected to one node in the other network.  In the first set of simulations, we used two Erd\"os-Renyi (ER) networks (Poisson degree distribution), while in the second set, we used two Barabasi-Albert (BA) or scale-free networks.    

\begin{figure}[]
\centering
\subfigure[]{\includegraphics[scale=0.22]{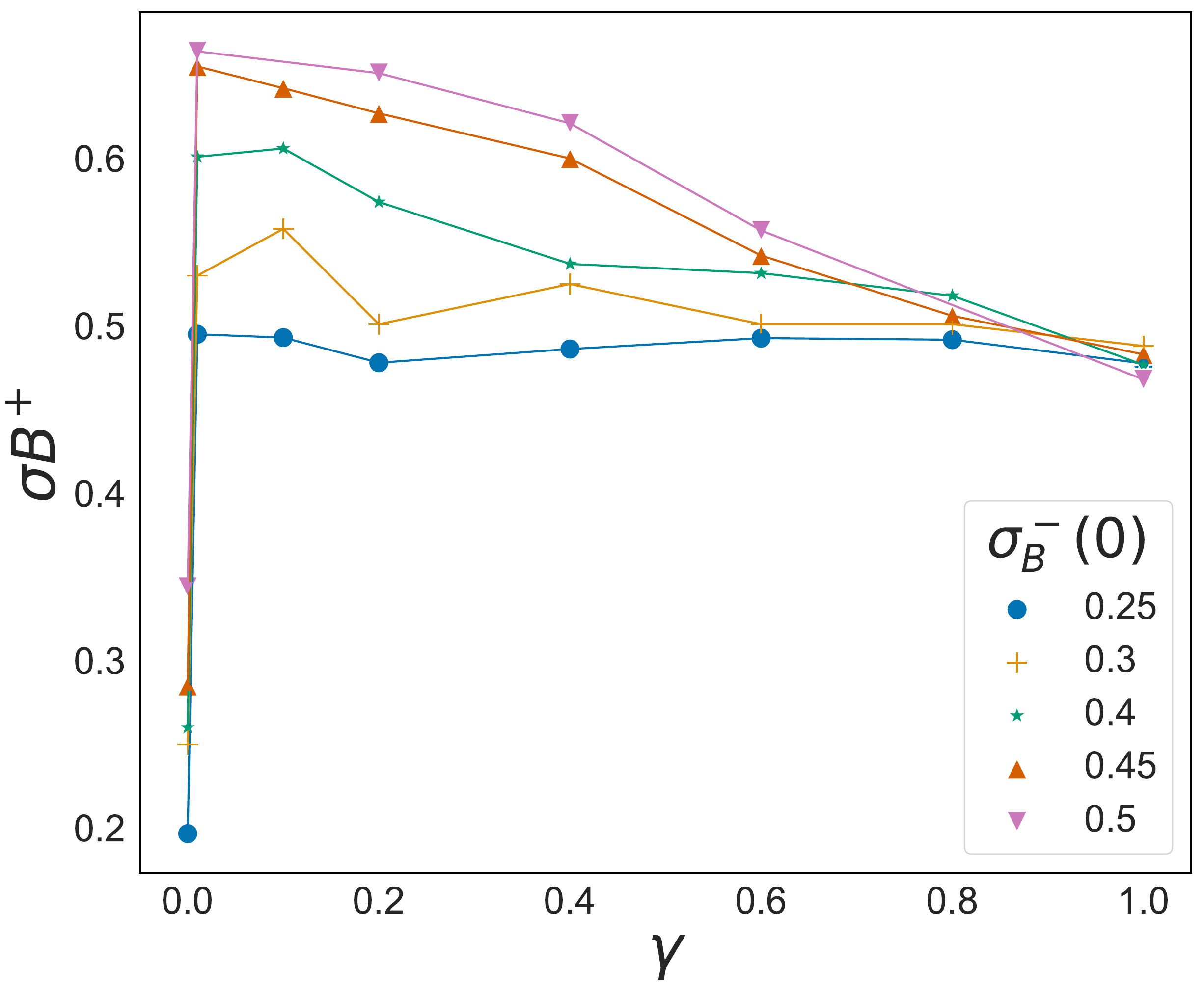}}
\subfigure[]{\includegraphics[scale=0.22]{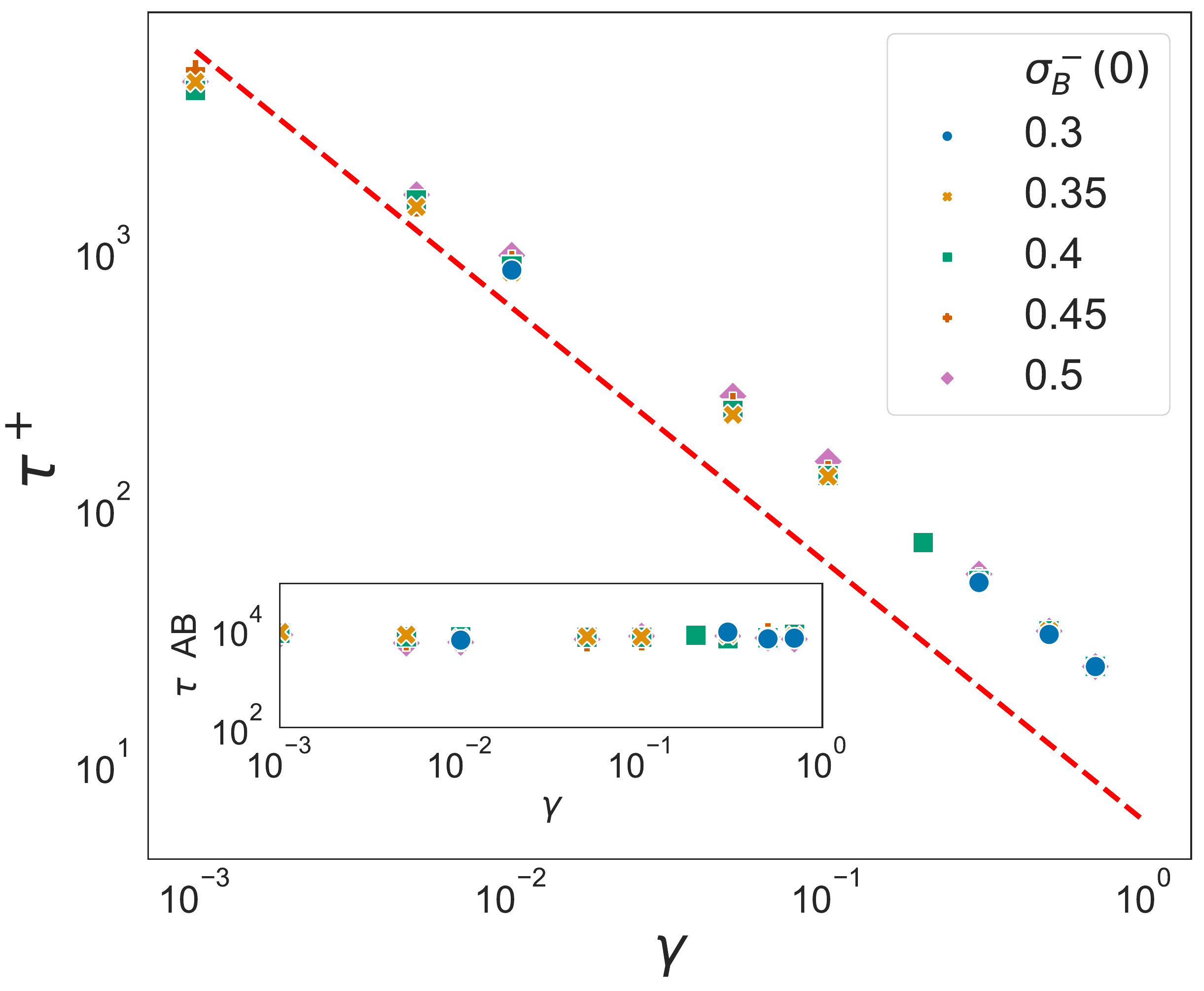}}
\subfigure[]{\includegraphics[scale=0.22]{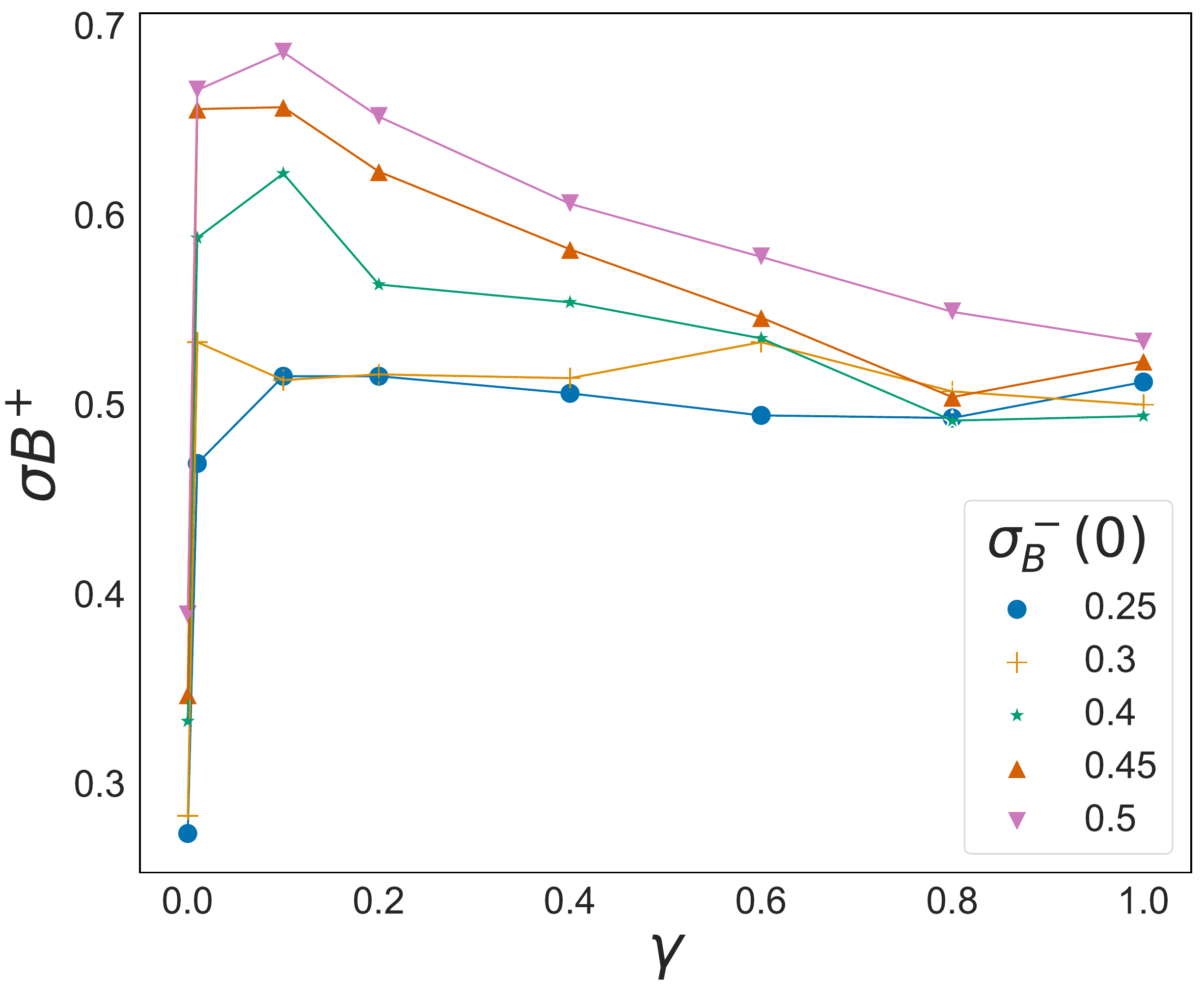}}
\subfigure[]{\includegraphics[scale=0.21]{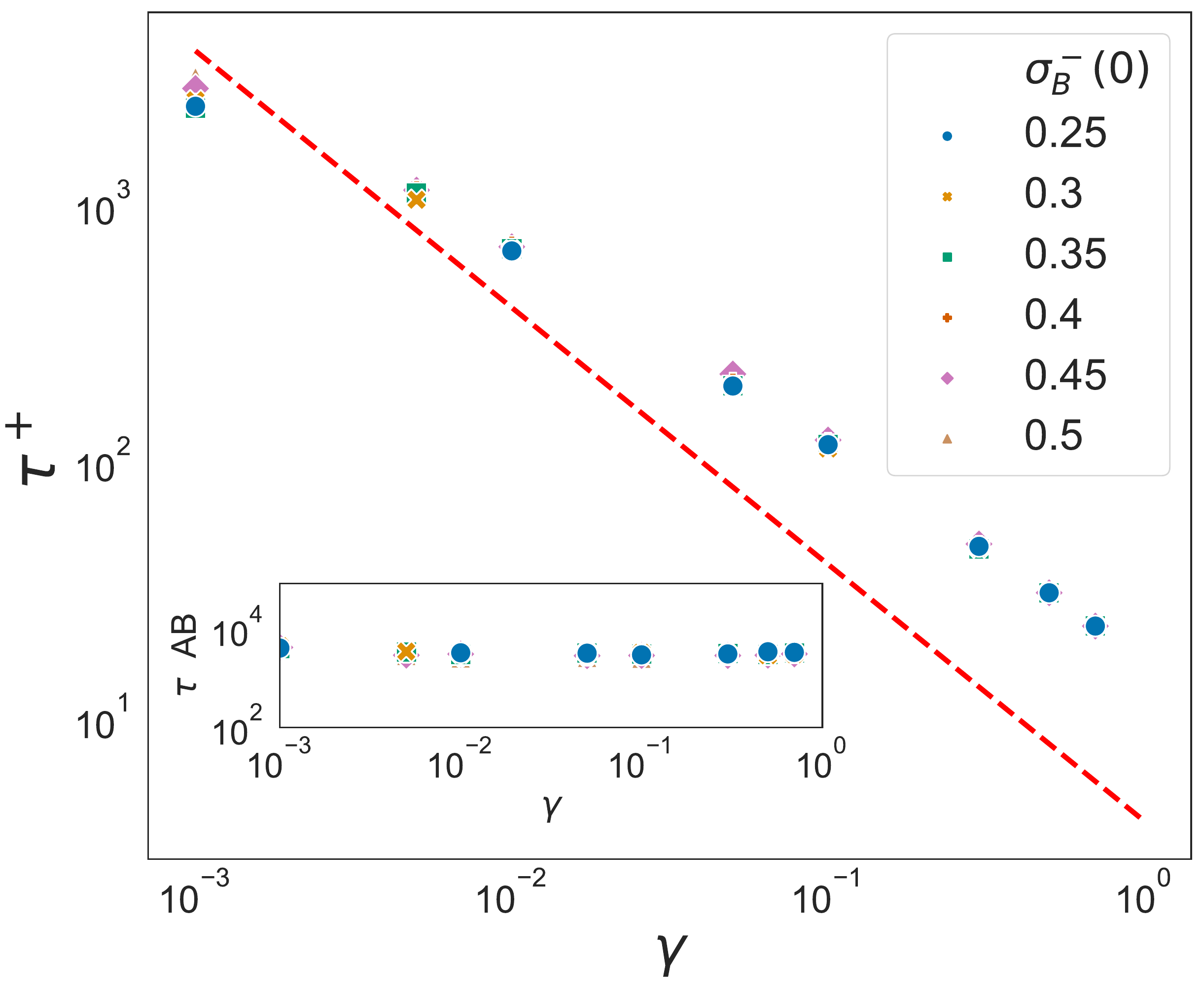}}
\caption{\small Monte-Carlo results of the simulated bi-layer voter model without the inclusion of bots. In (a) and (b), we have the average value of the final densities of tolerant individuals with opinion $B$ ($\sigma_{B}^+$) and the consensus time, both as a function of $\gamma$. The dotted red line represents the $\gamma^{-1}$ curve. In the inset figure, we have the consensus time for the Layer AB. Top panels (a) and (b) are for the synthetic Erdös-Renyi (ER) network, and the bottom panels (c) and (d) are the analysis for the Barabási-Albert (BA) network. The legends indicate the initial density of radical individuals with opinion B.}
\label{fig:impactRealNoBots}
\end{figure}

We notice that the only possible final state in the simulations is the fully ordered or consensus state, in which all agents have the same opinion and tolerance level, unlike in the MF analysis of the model without bots, where a stationary coexistence of both opinions is possible.  This is because fluctuations in finite-size networks make the system ultimately fall in an absorbing state of complete order, where the system is trapped and can no longer evolve, while MF equations are for infinite large systems and neglect fluctuations.  The results we present in this section correspond to average values over $500$ independent realizations of the dynamics for each initial condition.

In Figure~\ref{fig:impactRealNoBots} we show simulation results of the model without bots.  Top panels (a) and (b) correspond to ER networks, while bottom panels (c) and (d) correspond to BA networks.  In panels (a) and (c) we plot the average value of the final density of tolerant agents with opinion $B$, $\sigma_B^*$, as a function of $\gamma$, where we observe for both ER and BA networks a behavior that is similar to that found with the MF approach [see Fig.~\ref{fig:impact}(a)], that is, the smaller the $\gamma$, the larger the $\sigma_B^*$.  Panels (b) and (d) show the mean consensus time to the tolerant state ($\tau^+$) as a function of $\gamma$, where we can see the decay of $\tau^+$ with $\gamma$ that approximately follows a power law with an exponent close to $-1$ (dashed line), in close agreement with the MF approach [Fig.~\ref{fig:impact}(b)].  This confirms that the system reaches a tolerant consensus which takes a time that increases as the intolerant agents become more resilient, i.e., as $\gamma$ decreases.  However, as we mentioned before, the system ultimately reaches consensus by fluctuations, something not captured by the MF equations.  In the insets of  Fig.~\ref{fig:impactRealNoBots}(b) and (d), we plot the mean opinion consensus time $\tau_{AB}$, where we see that $\tau_{AB}$ is independent of $\gamma$ and of order $N=10^4$.  This is because the dynamic that leads to the final opinion consensus is that of the voter model between two symmetric states $A^+$ and $B^+$, which scales as $\tau_{AB} \sim N$, and does not depend on $\gamma$ because there are no intolerant agents.

\begin{figure}[]
\centering
\subfigure[]{\includegraphics[scale=0.22]{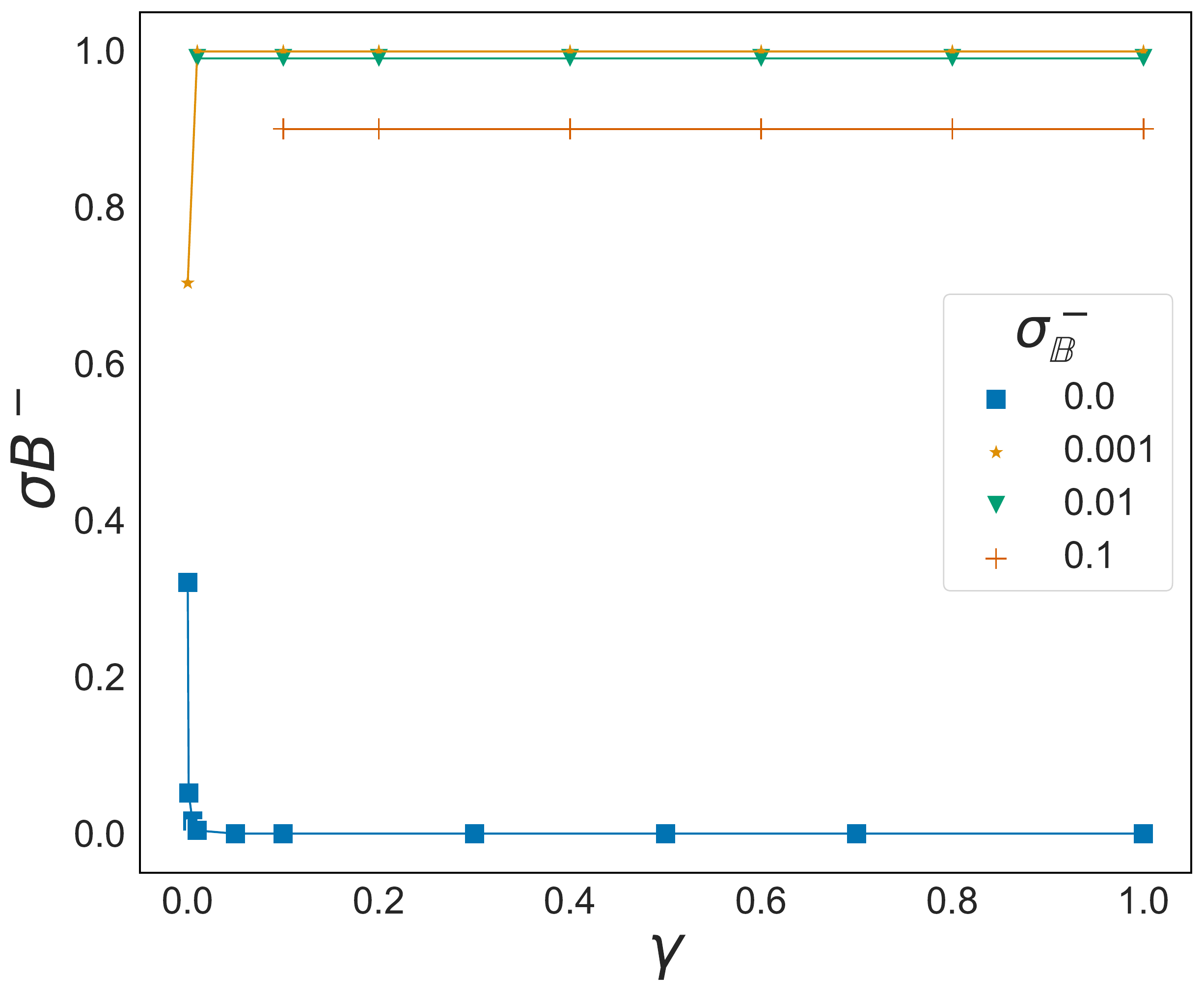}}
\subfigure[]{\includegraphics[scale=0.22]{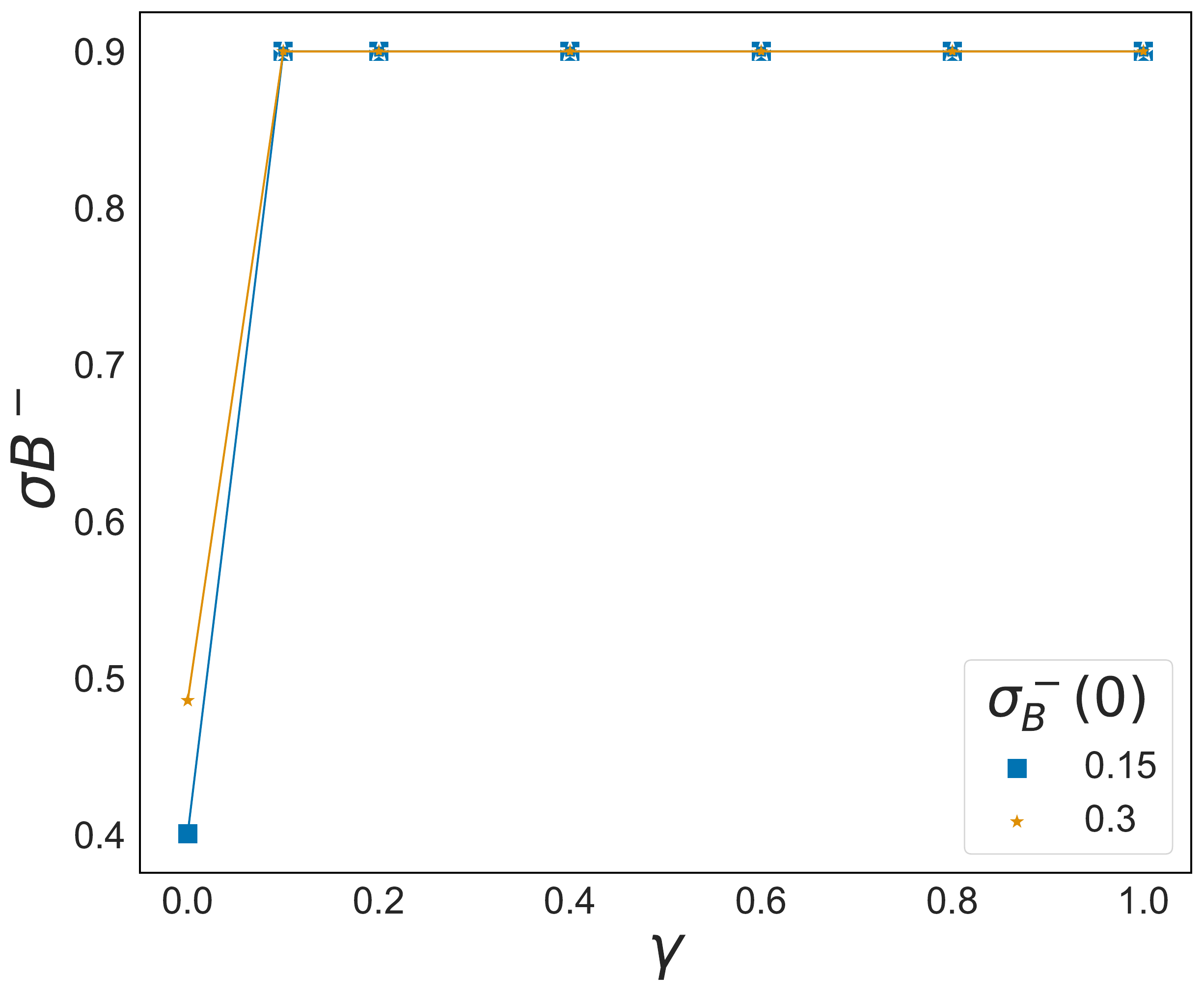}}
\subfigure[]{\includegraphics[scale=0.22]{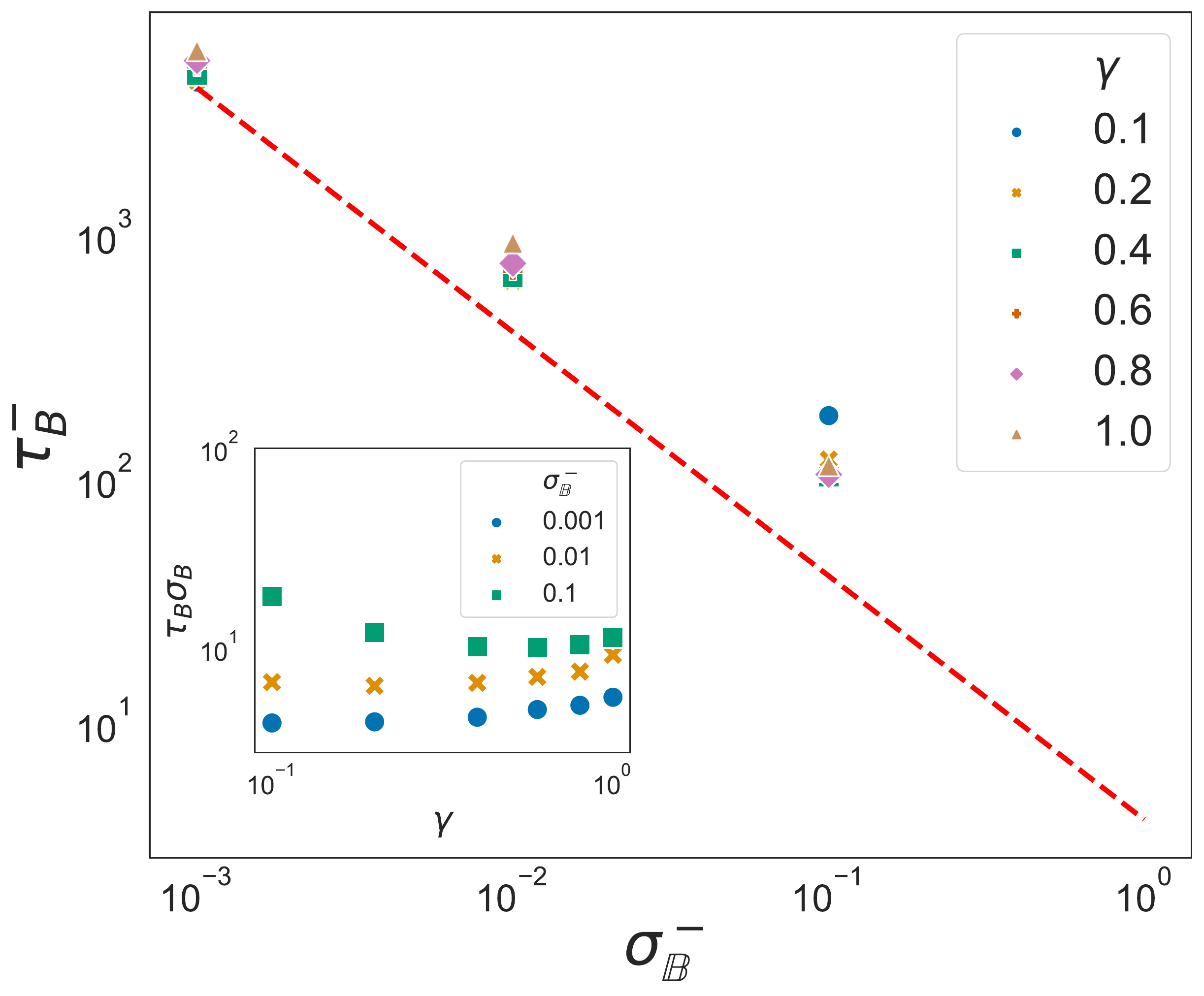}}
\subfigure[]{\includegraphics[scale=0.22]{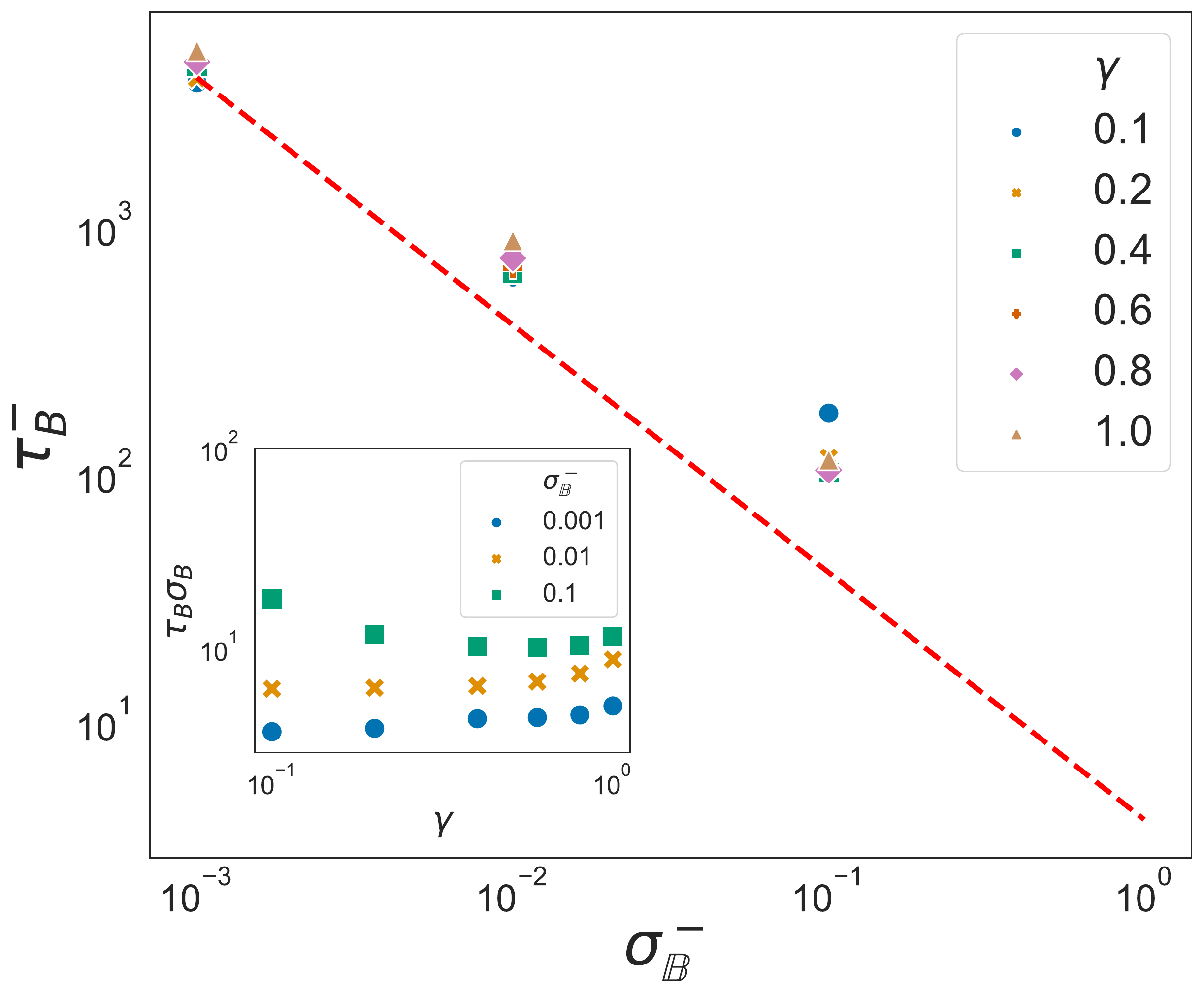}}
\caption{\small Monte Carlo simulation results of the bi-layer voter model with the inclusion of bots. Figures (a) and (b) show the final density of opinion $B$ intolerant agents ($\sigma_B^-$) as a function of $\gamma$ for different initial conditions; in (b) we fixed to 10\% the number of bots. Figures (c) and (d) show the average consensus time for the ER and BA networks, respectively; In the inset figures, we have the consensus time to reach the final density of B agents, in terms of the density of bots $\sigma_{\mathbb{B}}^{-}$. The dotted lines are the $1/\sigma_{\mathbb{B}}^{-}$ curve.}
\label{fig:impactRealBots}
\end{figure}

In Fig.~\ref{fig:impactRealBots} we show simulation results of the model with bots.  Panels (a) and (b) show the final density of opinion $B$ intolerant agents ($\sigma_B^-$) as a function of $\gamma$ for different initial conditions.  In agreement with MF results [Figs.~\ref{fig:impactB}(a) and (b)], the system always reaches a consensus of $B^-$ agents for $\gamma>0$ and $\sigma_{\mathbb{B}}^{-} > 0$, independent of the initial condition, while for $\gamma=0$ the final state consists of an absorbing configuration with a coexistence of $A^-$ and $B^-$ agents.  Panels (c) and (d) show the mean consensus time to the $B^-$ state ($\tau_B^-$) for ER and BA networks, respectively.  We observe that $\tau_B^-$ decays as power law with exponent close to $-1$ (dashed line), as predicted by the MF theory [Figure~\ref{fig:impactB}(c)].  That is, the consensus time increases as the fraction of bots decreases.  In the insets of panels (c) and (d) we see that $\tau_B^-$ does not change much with $\gamma$.  This is probably due to the fact that the values of $\gamma$ used in simulations were not small enough (simulation are computationally very costly for $\gamma<0.1$), possible hindering the power law behavior $\tau_B^- \sim \gamma^{-1}$ found in the MF approximation [Figure~\ref{fig:impactB}(d)].

\section{\label{sec:conclsuions} Summary and conclusions}

We proposed a voter model on a multiplex network to study the interplay between the dynamics of opinions ($A$ and $B$) and the tolerance (tolerant/intolerant) of individuals to accept others' opinions.  Intolerant agents are less likely to change opinion, and they become tolerant when they do.  We have also explored the effects of introducing a fraction of agents that play the role of bots, which are entities that never change opinion but can intentionally align other agents' opinions in a given direction.  We performed simulations on Erd\"os-Renyi and Barabási-Albert networks and studied the system using an MF approach. 
When there are no bots in the population, both opinion states are symmetric. The system is initially driven towards a state where all agents are tolerant, with fractions of $A$ and $B$--opinion agents that depend on the initial condition.  This consensus of tolerant agents happens because there is a bias of agents from the intolerant to the tolerant state.  After this first stage of tolerant consensus, there is a second stage where both opinions of tolerant agents evolve under the voter dynamics.  As this dynamics in a finite system is only driven by finite-size fluctuations, the fractions of voters of each opinion perform a symmetric random walk until a consensus in one opinion is eventually reached.  This final state of consensus is absorbing, as opinion and tolerance states can no longer evolve, unlike the initial tolerant consensus that is an active state where both opinions coexist.  The time to reach the initial tolerant consensus scales as $\tau^+ \sim \gamma^{-1}$, given that it is controlled by the rate $\gamma$ at which intolerant agents become tolerant.  Consequently, radical agents can slow down the dynamics towards consensus by a factor that diverges as they become more persistent in their opinions ($\gamma \to 0$).  The time to reach the final opinion consensus scales as in the voter model, $\tau_{AB} \sim N$, where $N$ is the number of agents.  Thus, the overall consensus time of the system is determined by $\gamma$ in the case of very intolerant agents ($\gamma \ll 1/N$) and by $N$ for very large systems ($N \gg \gamma^{-1}$).  

Adding in the population bots that hold opinion $B$ breaks the system's symmetry in both opinion and tolerance states, introducing a bias towards the intolerant opinion $B$ state.  This broken symmetry dramatically changes the model's outcome, where bots eventually impose their opinion to the rest of the system.  As bots behave as intolerant agents, the final (absorbing) consensus state consists of all intolerant agents with opinion $B$.  The consensus time has a non-trivial dependence on $\gamma$ and the fraction of bots $\sigma_{\mathbb{B}}^-$, where the first controls the time scale associated with the persistence of intolerant agents and the second controls the bias towards intolerant opinion $B$.  In the limiting case scenarios the consensus time is determined by the slowest of these two time scales, that is, $\tau_B^- \sim \gamma^{-1}$ for $\gamma \ll \sigma_{\mathbb{B}}^-$ and $\tau_B^- \sim 1/\sigma_{\mathbb{B}}^-$ for $\sigma_{\mathbb{B}}^- \ll \gamma$.  

The results described above mean that radical individuals who are resilient to change their minds can significantly impact the consensus of opinions, slowing down the overall opinion consensus process.  However, a striking consequence of the existence of radical or extremist individuals is that the entire population eventually becomes tolerant,  in a state having only moderate individuals of both opinions, which are more prone to change and reach consensus.  Therefore, the consensus of opinions in the model is a two-step process characterized by an initial extinction of extremists --who hinder opinion consensus-- and a later debate between moderate individuals that facilitate consensus.  Contrary to this result, bots can have the negative effect of preventing the state of tolerant consensus and leading the population to a state where every individual is an extremist of the opinion imposed by bots, which can be risky in democratic societies. 

It might be interesting to study an extension of the model where intolerant agents switch opinion with a probability that depends on its opinion $A$ or $B$, i.e., $\gamma_A$ and $\gamma_B$, respectively.  This could model a society where the level of individuals' tolerance depends on their opinion orientations, for instance, rightist or leftist.  Also, it would be worthwhile to explore a version of the model with a quote of free will by adding the possibility that agents switch opinion spontaneously, modeled as external noise.  These are variants of the model for future investigation.

\section*{Acknowledgements}

This research is supported by the Fundação de Amparo à Pesquisa do Estado de São Paulo (FAPESP) under Grant No.: 2015/50122-0 and the German Research Council (DFG-GRTK) Grant No.: 1740/2. D.A.V.O acknowledges the computational resources from the Center for Mathematical Sciences Applied to Industry (CeMEAI)  under Grant 2013/07375-0, and FAPESP Grants 2016/23698-1, 2018/24260-5, and 2019/26283-5. F.V. acknowledges financial support from Agencia Nacional de Promoci\'on Cien\'itfica y Tecnol\'ogica (Grant No. PICT 2016 Nro 201-0215). H.L.C.G. was funded by the research scholarship PCI-INPE, process 301113/2020-3. We thank Prof. Dr. Alessandro Vespignani, Dr. Dario Mazzilli, and PhD(c) Daniele Notarmuzi for useful comments and intellectual discussions.

\appendix

\renewcommand{\thesection}{Appendix \Alph{section}}

\renewcommand{\thetable}{\Alph{section}.\arabic{table}}

\section{\label{sec:transitions}Complement of the explicit transitions rules}

\begin{table}[]
	\centering
    \caption{\small Explicit transitions rules of the baseline bi-layer voter model without bots.}
		\begin{tabular}{|c | c c c | c c c |}
		\hline
		\multicolumn{7}{|c|}{\small{Bi-layer voter transitions without Bots}} \\ 
 		\hline \hline
 		
 	\multirow{3}{*}{\rotatebox[origin=c]{90}{ Layer $\pm$ \ }}
 	& $A^+ \ A^-$ & $\longrightarrow$ & $A^- \ A^-$ & $B^+ \ B^-$ & $\longrightarrow$ & $B^- \ B^-$ \\ 
 		& $A^- \ A^+$ & $\longrightarrow$ & $A^+ \ A^+$ & $B^- \ B^+$ & $\longrightarrow$ & $B^+ \ B^+$ \\ 
 		& $A^+ \ B^-$ & $\longrightarrow$ & $A^- \ B^-$ & $B^+ \ A^-$ & $\longrightarrow$ & $B^- \ A^-$ \\ 
 		& $A^- \ B^+$ & $\longrightarrow$ & $A^+ \ B^+$ & $B^- \ A^+$ & $\longrightarrow$ & $B^+ \ A^+$ \\ 
 		
	\hline
	 \multirow{3}{*}{\rotatebox[origin=c]{90}{Layer AB }} 	
		& $A^+ \ B^+$ & $\stackrel{}{\longrightarrow}$  & $B^+ \ B^+$ 
		& $B^+ \ A^+$ & $\stackrel{}{\longrightarrow}$  & $A^+ \ A^+$ \\ 
		& $A^+ \ B^-$ & $\stackrel{}{\longrightarrow}$  & $B^+ \ B^-$ 
		& $B^+ \ A^-$ & $\stackrel{}{\longrightarrow}$  & $A^+ \ A^-$ \\ 
	& {$A^- \ B^+$} & $\stackrel{{\gamma}}{\longrightarrow}$  & $B^+ \ B^+$ & {$B^- \ A^+$} & $\stackrel{{\gamma}}{\longrightarrow}$  & $A^+ \ A^+$ \\
	
	& {$A^- \ B^-$} & $\stackrel{{\gamma}}{\longrightarrow}$  & $B^+ \ B^-$
	& {$B^- \ A^-$} & $\stackrel{{\gamma}}{\longrightarrow}$  & $A^+ \ A^-$ \\
	
	\hline
	\end{tabular}
	\label{table-no-bots}
\end{table}
\begin{table}[]
	\centering
	\caption{\small Explicit transition rules of the bi-layer voter model including bots.}
		\begin{tabular}{|c | c c c | c c c |}
		\hline
		\multicolumn{7}{|c|}{\small{Bi-layer voter transitions including Bots}} \\ 
 		\hline \hline
 	\multirow{3}{*}{\rotatebox[origin=c]{90}{ Layer $\pm$ \ }}
 	    & {$A^+ \ A^-$} & $\longrightarrow$ & $A^- \ A^-$ & {$B^+ \ (B^-+\mathbb{B}^-)$} & $\longrightarrow$ & $B^- \ (B^-+\mathbb{B}^-)$\\ 
 		& {$A^- \ A^+$} & $\longrightarrow$ & $A^+ \ A^+$ &  
 		 {$ (B^-+\mathbb{B}^-)\ B^+$} & $\longrightarrow$ & $(B^++\mathbb{B}^-) \ B^+$ \\ 
 		& {$A^+ \ (B^-+\mathbb{B}^-)$} & $\longrightarrow$ & $A^- \ (B^-+\mathbb{B}^-)$ & 
 		 {$B^+ \ A^-$} & $\longrightarrow$ & $B^- \ A^-$ \\ 
 		& {$A^- \ B^+$} & $\longrightarrow$ & $A^+ \ B^+$ & {$(B^-+\mathbb{B}^-) \ A^+$} & $\longrightarrow$ & $(B^++\mathbb{B}^-) \ A^+$ \\ 
	\hline
	 \multirow{3}{*}{\rotatebox[origin=c]{90}{Layer AB }} 
		& {$A^+ \ B^+$} & $\stackrel{}{\longrightarrow}$  & $B^+ \ B^+$ 
		& {$B^+ \ A^+$} & $\stackrel{}{\longrightarrow}$  & $A^+ \ A^+$ \\ 
		& {$A^+ \ (B^-+\mathbb{B}^-)$} & $\stackrel{}{\longrightarrow}$  & $B^+ \ (B^-+\mathbb{B}^-)$ 
		& {$B^+ \ A^-$} & $\stackrel{}{\longrightarrow}$  & $A^+ \ A^-$ \\ 	
	    & {{$A^- \ B^+$}} & $\stackrel{{\gamma}}{\longrightarrow}$  & $B^+ \ B^+$ & {{$(B^-+\mathbb{B}^-) \ A^+$}} & $\stackrel{{\gamma}}{\longrightarrow}$  & $(A^++\mathbb{B}^-) \ A^+$ \\
	    & {{$A^- \ (B^-+\mathbb{B}^-)$}} & $\stackrel{{\gamma}}{\longrightarrow}$  & $B^+ \ (B^-+\mathbb{B}^-)$
	    & {{$(B^-+\mathbb{B}^-) \ A^-$}} & $\stackrel{{\gamma}}{\longrightarrow}$  & $(A^-+\mathbb{B}^-) \ A^-$ \\
	\hline
	\end{tabular}
	\label{table-bots}
\end{table}

In this section we explicitly write all transitions between opinion and tolerance states of agents in the model without bots (table \ref{table-no-bots}) and with bots (table \ref{table-bots}).  The notation $A^+, A^-, B^+$ and $B^-$ correspond to states of agents with opinion and tolerance $A$ and $+$, $A$ and $-$, $B$ and $+$, and $B$ and $-$, respectively.  In a single time step $\Delta t = 1/N$ of the dynamics, one node is chosen at random.  Then this nodes copies the tolerance of a random neighbor in the $\pm$--layer, and the opinion of a random neighbor in the $AB$--layer.  In tables \ref{table-no-bots} and \ref{table-bots}, the states on the left and right of a given pair correspond, respectively, to the focal agent --who changes state-- and the random neighbor on the corresponding layer.  Only situations that lead to a state change are included in the tables.

\bibliographystyle{amsplain}

\providecommand{\bysame}{\leavevmode\hbox to3em{\hrulefill}\thinspace}
\providecommand{\MR}{\relax\ifhmode\unskip\space\fi MR }
\providecommand{\MRhref}[2]{%
  \href{http://www.ams.org/mathscinet-getitem?mr=#1}{#2}
}
\providecommand{\href}[2]{#2}

\end{document}